\pdfoutput=1

\documentclass[onecolumn]{emulateapj}

\usepackage{amsmath}
\usepackage{graphicx}
\usepackage{color}
\usepackage[colorlinks]{hyperref}
\hypersetup{
    colorlinks,	
    citecolor=blue,
}

\newcommand{\be}{\begin{eqnarray}}
\newcommand{\ee}{\end{eqnarray}}

\shorttitle{Testing the Kerr hypothesis with a disk with finite thickness}
\shortauthors{Abdikamalov et al.}

\begin{document}

\title{Testing the Kerr black hole hypothesis using X-ray reflection spectroscopy and a thin disk model with finite thickness}

\author{Askar~B.~Abdikamalov\altaffilmark{1,2}, Dimitry~Ayzenberg\altaffilmark{1}, Cosimo~Bambi\altaffilmark{1,\dag}, Thomas~Dauser\altaffilmark{3}, Javier~A.~Garc\'ia\altaffilmark{4,3}, Sourabh~Nampalliwar\altaffilmark{5}, Ashutosh~Tripathi\altaffilmark{1}, and Menglei~Zhou\altaffilmark{1}}

\altaffiltext{1}{Center for Field Theory and Particle Physics and Department of Physics, 
Fudan University, 200438 Shanghai, China. \email[\dag E-mail: ]{bambi@fudan.edu.cn}}
\altaffiltext{2}{Ulugh Beg Astronomical Institute, Tashkent 100052, Uzbekistan}
\altaffiltext{3}{Remeis Observatory \& ECAP, Universit\"at Erlangen-N\"urnberg, 96049 Bamberg, Germany}
\altaffiltext{4}{Cahill Center for Astronomy and Astrophysics, California Institute of Technology, Pasadena, CA 91125, USA}
\altaffiltext{5}{Theoretical Astrophysics, Eberhard-Karls Universit\"at T\"ubingen, 72076 T\"ubingen, Germany}

\begin{abstract}
X-ray reflection spectroscopy is a powerful tool for probing the strong gravity region of black holes and can be used for testing general relativity in the strong field regime. Simplifications of the available relativistic reflection models limit the capability of performing accurate measurements of the properties of black holes. In this paper, we present an extension of the model {\sc relxill\_nk} in which the accretion disk has a finite thickness rather than being infinitesimally thin. We employ the accretion disk geometry proposed by~\citet{2018ApJ...855..120T} and we construct relativistic reflection models for different values of the mass accretion rate of the black hole. We apply the new model to high quality \textsl{Suzaku} data of the X-ray binary GRS~1915+105 to explore the impact of the thickness of the disk on tests of the Kerr metric. 
\end{abstract}



\section{Introduction}

Einstein's theory of general relativity is a pillar of modern physics and in agreement with all the available observational tests~\citep{2014LRR....17....4W}. However, the theory has been primarily tested in weak gravitational fields, while its predictions in the strong field regime have only recently being put to test. Astrophysical black holes are ideal laboratories for testing general relativity in the strong field regime and a number of theoretical reasonings point to the possibility that the spacetime metric around these objects can present macroscopic deviations from the predictions of Einstein's gravity~\citep[see, for instance,][]{e3,e4,e5}.

In 4-dimensional general relativity, uncharged black holes are relatively simple systems. They are described by the Kerr solution~\citep{k1} and are completely specified by only two parameters, representing, respectively, the mass $M$ and the spin angular momentum $J$ of the black hole. This is the well-known conclusion of the no-hair theorems, and it holds under specific assumptions~\citep{k2,k3,k4}. It is also quite remarkable that the spacetime metric around an astrophysical black hole formed from the complete collapse of a progenitor body should be well approximated by the simple Kerr solution. For example, the presence of an accretion disk or of a nearby star has a very small impact on the near horizon metric and can normally be ignored~\citep{d1,d2}. The search for possible deviations from the Kerr geometry in the strong gravity region of an astrophysical black hole can thus be a tool to constrain and find new physics.

The Kerr black hole hypothesis can be tested by studying the properties of the electromagnetic radiation emitted by material orbiting a black hole~\citep{review,2016CQGra..33l4001J,2018GReGr..50..100K,2019PhRvD..99j4031Z}. Among all the electromagnetic techniques for testing the near horizon region of black holes, X-ray reflection spectroscopy~\citep{1989MNRAS.238..729F,2006ApJ...652.1028B,2014SSRv..183..277R} is the most mature one and the only one that can currently provide quantitative constraints on the black hole strong gravity region~\citep[see, for instance,][]{2018PhRvL.120e1101C,2019ApJ...875...56T,2019ApJ...874..135T,2019ApJ...884..147Z}. Like any astrophysical measurement, even for X-ray reflection spectroscopy it is crucial to have a sufficiently sophisticated astrophysical model in order to limit the modeling systematic uncertainties.

X-ray reflection spectroscopy refers to the analysis of the features of the reflection spectrum of accretion disks. Our system is a central black hole accreting from a geometrically thin and optically thick disk, with the inner edge of the disk at the innermost stable circular orbit (ISCO). Similar disks are thought to form when the source is in the thermal state with an accretion luminosity between a few percent and about 30\% of its Eddington limit~\citep{2006ApJ...652..518M,2010MNRAS.408..752P,2010ApJ...718L.117S}. The gas of the accretion disk is in local thermal equilibrium and at any point on the surface of the disk the emission is like that of a blackbody. The spectrum of the whole disk is a multi-temperature blackbody-like spectrum because the temperature increases as the gas falls into the gravitational well of the black hole~\citep{1974ApJ...191..499P,1997ApJ...482L.155Z}. The thermal emission of the accretion disk is normally peaking in the soft X-ray band (0.1-1~keV) for stellar-mass black holes and in the optical/UV band (1-100~eV) for supermassive ones, as the disk temperature scales as $M^{-0.25}$~\citep{1997ApJ...482L.155Z}. The ``corona'' is some hotter ($\sim 100$~keV), usually compact and optically thin, gas near the black hole. Thermal photons from the disk can inverse Compton scatter off free electrons in the corona, producing a power-law component with an exponential cut-off in the X-ray spectrum of the black hole~\citep{1979Natur.279..506S}. The Comptonized photons can illuminate the accretion disk, producing the reflection component~\citep{1991MNRAS.249..352G,2005MNRAS.358..211R,2013ApJ...768..146G}. The latter is characterized by fluorescent emission lines below 8~keV, notably the iron K$\alpha$ complex at 6.4-6.79~keV depending on the ionization of iron ions, and the so-called Compton hump peaking at 20-30~keV.

A relativistic reflection model relies on a model to calculate the reflection spectrum at every emission point on the disk (assuming Einstein's Equivalence Principle holds, these calculations only involve atomic physics) as well as on a disk-corona model\footnote{For ``disk-corona model'' here we mean a model for the description of the accretion disk and the assumptions on the coronal geometry to describe the disk's intensity profile.} and a spacetime metric, which are both necessary to calculate the reflection spectrum at the detection point far from the source. All these pieces have a number of parameters and a variation in the value of these model parameters can have an impact on the predicted reflection spectrum of an accreting black hole. Fitting observational data with the theoretical model, we can infer the value of the model parameters and thus the properties of the system. If we employ a spacetime metric with some parameters quantifying deviations from the Kerr spacetime, we can attempt to constrain possible deviations from the Kerr metric by fitting X-ray data of some reflection-dominated source with our model.

{\sc relxill\_nk} is a relativistic reflection model to test the Kerr black hole hypothesis~\citep{2017ApJ...842...76B,2019ApJ...878...91A}. It is an extension of the {\sc relxill} package~\citep{2013MNRAS.430.1694D,2013ApJ...768..146G,2014ApJ...782...76G} to non-Kerr spacetimes. As in {\sc relxill}, in {\sc relxill\_nk} the reflection spectrum in the rest-frame of the disk is modeled by {\sc xillver}, the accreting matter is described by an infinitesimally thin Novikov-Thorne disk~\citep{1973blho.conf..343N,1974ApJ...191..499P}, and the disk's intensity profile is either described by a broken power-law or is the profile generated by a corona with lamppost geometry. {\sc relxill\_nk} differs from {\sc relxill} only in the spacetime metric. The main version of {\sc relxill\_nk} employs the Johannsen metric~\citep{jj}, which is not an exact solution of some specific gravity model but a parametric black hole spacetime. The Johannsen metric has an infinite number of ``deformation parameters'' that quantify deviations from the Kerr background. With the spirit of a null-experiment, we can fit the reflection spectrum of a source with {\sc relxill\_nk}, determine the values of the deformation parameters, and thus verify if they are consistent with the hypothesis that the metric around the source is described by the Kerr solution as required by general relativity. As it has been constructed, {\sc relxill\_nk} can easily employ any stationary, axisymmetric, and asymptotically flat metric in analytic form~\citep[see, e.g.,][]{2018PhRvD..98b4007Z,2019arXiv190312119N,2019arXiv190805177Z,2019arXiv191203868T}.

Like in any astrophysical measurement, even for the tests of the Kerr metric with {\sc relxill\_nk}, it is crucial to limit the systematic uncertainties. Otherwise, in the presence of high quality data, we could obtain precise but inaccurate measurements of the spacetime metric around an accreting black hole and our analysis may find deviations from the Kerr solution that, actually, are due to systematic uncertainties. Among all the systematic uncertainties, modeling uncertainties are normally the dominant ones. {\sc relxill\_nk} has a number of modeling uncertainties, ranging from simplifications in the non-relativistic reflection model and in the disk-corona model to relativistic effects not taken into account~\citep[see, for instance, the discussion in][]{2019PhRvD..99l3007L,2020PhRvD.101d3010Z}.

All the available relativistic reflection models assume that the black hole accretion disk is geometrically thin and that there is no emission of radiation inside the inner edge of the disk. For example, if we apply these models to sources accreting near their Eddington limit, the spin parameter can be easily overestimated~\citep{2019arXiv191106605R,2020MNRAS.491..417R}. Moreover, the accretion disk is always approximated as infinitesimally thin. For a real accretion disk, we should expect that the disk has a finite thickness and that the latter increases as the mass accretion rate increases. Employing a model with an infinitesimally thin accretion disk inevitably leads to modeling bias in the final measurements of some model parameters. The impact of such a simplification has been ignored for a long time and only recently \citet{2018ApJ...855..120T} have presented a relativistic reflection model in which the accretion disk has a finite thickness. In the present paper, we implement the accretion disk model of \citet{2018ApJ...855..120T} into {\sc rellxil\_nk} as a step of our program of developing relativistic reflection models for testing the Kerr black hole hypothesis in order to try to create a tool for precision tests of general relativity in the strong field regime. The implementation of a disk of finite thickness in {\sc rellxil\_nk} can be useful to analyze those sources with thicker accretion disks and providing more precise measurements of the deformation parameters of the spacetime.

The paper is organized as follows. In Section~\ref{sec:disk}, we review the accretion disk geometry proposed in~\citet{2018ApJ...855..120T} and, in Section~\ref{sec:trf}, we employ such a disk geometry in {\sc relxill\_nk}. In Section~\ref{sec:grs1915}, we use the new model with a disk of finite thickness to analyze a \textsl{Suzaku} observation of the X-ray binary GRS~1915+105 and to explore the impact of the disk thickness in our tests of the Kerr metric. This \textsl{Suzaku} observation of GRS~1915+105 was studied in~\citet{2019ApJ...884..147Z} and currently provides one of the most precise measurements of the deformation parameters of the spacetime, so it is presumably quite sensitive to systematic uncertainties. Summary and conclusions are reported in Section~\ref{sec:conclusions}. In Appendix~\ref{app:metric}, we briefly review the Johannsen metric and its black hole parameter space. Throughout the paper, we use units in which $G_{\rm N} = c = 1$ and a metric with signature $(-+++)$.

\vspace{0.3cm}


\section{Accretion disk}\label{sec:disk}

We consider the accretion disk geometry proposed in~\citet{2018ApJ...855..120T}. We assume that the accretion disk mid-plane lies in the $\theta=\pi/2$ plane and is a radiation-pressure dominated, geometrically thin, and optically thick disk with a pressure scale height $H$ defined as~\citep{1973A&A....24..337S}
\be\label{eq:Hdef}
H=\frac{3}{2}\frac{1}{\eta}\left(\frac{\dot{M}}{\dot{M}_{\rm Edd}}\right)\left[1-\sqrt{\frac{r_{\rm ISCO}}{\rho}}\right],
\ee
where $\rho=r\sin\theta$ is the pseudo-cylindrical radius, $ \dot{M} / \dot{M}_{\rm Edd}$ is the Eddington-scaled mass accretion rate, and $r_{\rm ISCO}$ is the ISCO radius, which is also the inner-edge of the disk. The radiative efficiency is $\eta=1-E_{\rm ISCO}$, where $E_{\rm ISCO}$ is the specific energy of a test-particle in the mid-plane at $r_{\rm ISCO}$. We assume that the surface of the disk is determined by the half-thickness $z(\rho)=2H$ and that the disk rotates cylindrically ($\dot{\theta}=0$), which means that all matter at some pseudo-cylindrical radius $\rho$ in the disk will have the same orbital velocity as the material at the same cylindrical radius in the equatorial plane ($\theta=\pi/2$). Note that both $\eta$ and $r_{\rm ISCO}$ are functions of the spacetime metric. Therefore, for given spacetime parameters, the Eddington ratio $\dot{M} / \dot{M}_{\rm Edd}$ can be used as the disk thickness parameter, since the geometric thickness increases as we increase $\dot{M} / \dot{M}_{\rm Edd}$. In this paper, for the sake of simplicity, we will always assume that the spacetime geometry is described by the simplest version of the Johannsen metric with the only possible non-vanishing deformation parameter $\alpha_{13}$, while all other deformation parameters will be set to zero~\citep{jj}; the expression of such a metric with some basic properties is reported in Appendix~\ref{app:metric}. Figure~\ref{fig:disk} illustrates our disks of finite thickness around Johannsen black holes with different values of spin and deformation parameters. We note that the pressure scale height $H$ in Eq.~(\ref{eq:Hdef}) is derived from a Newtonian model~\citep{1973A&A....24..337S}. The actual pressure scale height must thus deviate from ours as we approach the black hole, affecting both the surface shape and the thickness of the disk. We can expect that the actual thickness of the disk is smaller for the same $\dot{M} / \dot{M}_{\rm Edd}$ and thus our model may overestimate a bit the effect; the different surface shape affects the actual value of the emission angle and, in turn, the Doppler boosting, which has an impact on the shape of the reflection spectrum. However, here the spirit is to extend our disk model from infinitesimally thin to finite thickness, as well as to get a crude estimate of the impact of the disk thickness on the measurement of the model parameters. A relativistic disk model would predict a somewhat different pressure scale height $H$, but to keep the analysis simple and facilitate the comparison with \citet{2018ApJ...855..120T} we use their model.

\begin{figure*}[t]
\begin{center}
\includegraphics[width=1.10\textwidth,trim={3.5cm 0cm 0cm 1.0cm},clip]{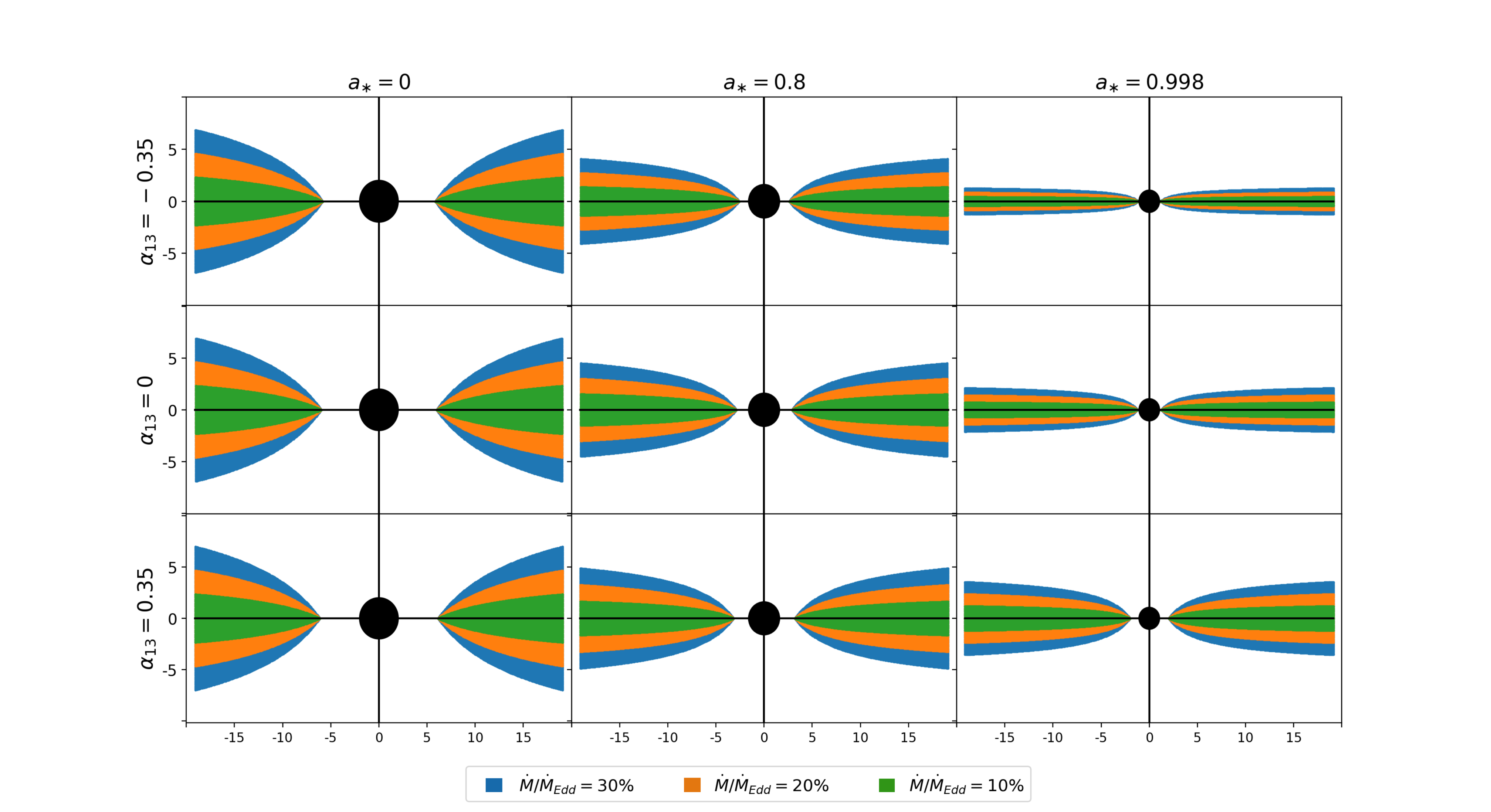}
\end{center}
\vspace{-0.2cm}
\caption{Accretion disk profiles for $\dot{M} / \dot{M}_{\rm Edd} = 0.1$, 0.2, and 0.3 in the case of $a_* = 0$, 0.8, and 0.998 and $\alpha_{13} = -0.35$, 0, and 0.35. $x$- and $y$-axes in units $M = 1$. Figure following \citet{2018ApJ...868..109T}.
\label{fig:disk}}
\end{figure*}

Since the spacetime is stationary and axisymmetric there are two Killing vectors, namely, a timelike and an azimuthal. Therefore, there are two conserved quantities: the specific energy $E$ and the $z$-component of the specific angular momentum $L_{z}$. The system is fully determined by imposing that the gas follows nearly-geodesic equatorial circular orbits~\citep{1972ApJ...178..347B}.

By definition, we can write
\begin{align}
\dot t =& -\frac{Eg_{\phi\phi}+L_{z}g_{t\phi}}{g_{tt}g_{\phi\phi}-g_{t\phi}^{2}}, \label{eq:dott}
\\
\dot\phi =& \frac{Eg_{t\phi}+L_{z}g_{tt}}{g_{tt}g_{\phi\phi}-g_{t\phi}^{2}}, \label{eq:dotphi}
\end{align}
where the overhead dot is a derivative with respect to the affine parameter (proper time for a massive particle). Employing Eqs.~(\ref{eq:dott}) and (\ref{eq:dotphi}) in the normalization condition for the 4-velocity of massive particles $u^{a}u_{a}=-1$, we get
\begin{equation}
g_{rr}\dot r^{2}+g_{\theta\theta}\dot\theta^{2}=V_{\text{eff}}(r,\theta;E,L_{z}),
\end{equation}
where the effective potential is
\begin{equation}
V_{\text{eff}}= -1 - \frac{E^{2}g_{\phi\phi}+2EL_{z}g_{t\phi}+L_{z}^{2}g_{tt}}{g_{tt}g_{\phi\phi} - g_{t\phi}^{2}}\, ,\label{eq:Veff}
\end{equation}
and the 4-velocity is $u^{a}=(\dot t, \dot r, \dot\theta, \dot\phi)$.

The explicit expressions for the energy and the angular momentum can be obtained when we impose equatorial circular orbits. The circularity condition is equivalent to require $V_{\text{eff}}=0$ and $\partial V_{\text{eff}}/\partial r=0$. If we solve for $E$ and $L_{z}$, we find
\begin{align}
E=&-\frac{g_{tt}+g_{t\phi}\Omega}{\sqrt{-(g_{tt}+2g_{t\phi}\Omega+g_{\phi\phi}\Omega^{2})}},\label{eq:E}
\\
L_{z}=&\frac{g_{t\phi}+g_{\phi\phi}\Omega}{\sqrt{-(g_{tt}+2g_{t\phi}\Omega+g_{\phi\phi}\Omega^{2})}},\label{eq:Lz}
\end{align}
where the angular velocity of equatorial circular geodesics is
\begin{equation}
\Omega=\frac{d\phi}{dt}=\frac{-g_{t\phi,r}\pm\sqrt{(g_{t\phi,r})^{2}-g_{tt,r}g_{\phi\phi,r}}}{g_{\phi\phi,r}}.\label{eq:angvel}
\end{equation}
With $\Omega$, we can write $\dot{t}$ from $u^{a}u_{a}=-1$ and considering that for equatorial circular orbits we have $u^{a}=(\dot t, 0, 0, \dot\phi)=(1, 0, 0, \Omega) \dot t $,
\begin{equation}
\dot t = \frac{1}{\sqrt{-(g_{tt}+2g_{t\phi}\Omega+g_{\phi\phi}\Omega^{2})}}.\label{eq:tdot}
\end{equation}

The ISCO radius can be calculated by substituting Eqs.~(\ref{eq:E}) and (\ref{eq:Lz}) into Eq.~(\ref{eq:Veff}) and solving $\partial^{2}V_{\text{eff}}/\partial r^{2}=0$ for $r$. 

\vspace{0.3cm}


\section{Transfer function}\label{sec:trf}

{\sc rellxil\_nk} employs the formalism of the transfer function for geometrically thin and optically thick accretion disks~\citep{1975ApJ...202..788C, 1995CoPhC..88..109S, 2010MNRAS.409.1534D}. The observed reflection spectrum is the sum of the observed specific intensities $I_{\rm o}(\nu_{\rm o})$ at frequency $\nu_{\rm o}$ from all parts of the disk. We can perform this sum by projecting the accretion disk onto a plane perpendicular to the line of sight of the observer, which corresponds to the observer's sky~\citep{1975ApJ...202..788C}.

The observer is located at spatial infinity $(r=+\infty)$ with inclination angle $\iota$ between the normal to the disk and the line of sight of the distant observer. We use Cartesian coordinates $(\alpha,\beta)$ on the observer's plane. In terms of the photon momentum, the celestial coordinates can be written as
\begin{equation}
\alpha=\lim_{r\rightarrow\infty}\frac{-rp^{(\phi)}}{p^{(t)}}, \quad
\beta=\lim_{r\rightarrow\infty}\frac{rp^{(\theta)}}{p^{(t)}}, \label{eq:celcoords}
\end{equation}
where $p^{(a)}$s are the components of the 4-momentum of the photon with respect to a locally non-rotating reference frame~\citep{1972ApJ...178..347B} and are related to $p^{a}$s through a coordinate transformation (e.g.~$p^{\phi}=p^{(\phi)}/\sin\iota$)\footnote{Note that this is the photon momentum of the incoming photon. There is a minus sign in the expression of $\alpha$ because $\alpha$ and $\beta$ are on the observer's screen, so the coordinates along that axis are mirrored, e.g., if a photon leaves the disk with positive $p^{\phi}$, i.e. moving to the left from the black hole's perspective, the photon will arrive on the observer's screen on the right.}. The celestial coordinates $(\alpha,\beta)$ and the solid angle on the observer's sky are related to each other through $d\alpha d\beta=D^{2}d\Omega$, where $D$ is the distance between the black hole and the observer~\citep{1975ApJ...202..788C}.

Using Liouville's theorem~\citep{1966AnPhy..37..487L}, that states $I_{\nu}/\nu^{3}=\text{const}.$, we can obtain the specific intensity as seen by the observer. The observed flux of an accretion disk can then be written as
\begin{equation}
F_{\rm o}(\nu_{\rm o})= \frac{1}{D^2}
\int g^{3}I_{\nu_{\rm e}}\left(r_{\rm e},\theta_{\rm e}\right)d\alpha d\beta,
\end{equation}
where $I_ {\nu_{\rm e}} (r_{\rm e},\theta_{\rm e})$ is the local specific intensity, $r_{\rm e}$ is the emission radius, $\theta_{\rm e}$ is the photon emission angle in the rest-frame of the gas, $\nu_{\rm e}$ is the photon frequency in the rest-frame of the gas, and $g$ is the redshift factor
\begin{equation}
g=\frac{\nu_{\rm o}}{\nu_{\rm e}}=\frac{(p_{a}u^{a})_{\rm o}}{(p_{b}u^{b})_{\rm e}}.\label{eq:redshift}
\end{equation}
Here $p^{a}$ is the 4-momentum of a photon, and $u^{a}_{\rm o}$ and $u^{a}_{\rm e}$ are the 4-velocities of the distant observer and the particles of the gas, respectively. The photon's 4-momentum is $p_{a}=(-E^{\gamma},p_{r},p_{\theta},L_{z}^{\gamma})$ and the observer is treated as static, $u_{\rm o}^{a}=(1,0,0,0)$. As we mentioned in Section~\ref{sec:disk}, the 4-velocity of the orbiting material in the accretion disk is $u_{\rm e}^{a}=\dot{t}(1,0,0,\Omega)$, where $\dot t$ is given by Eq.~(\ref{eq:tdot}) and $\Omega$ is given by Eq.~(\ref{eq:angvel}). If we plug Eq.~(\ref{eq:angvel}) and Eq.~(\ref{eq:tdot}) into Eq.~(\ref{eq:redshift}),  the redshift factor becomes
\begin{equation}
g=\frac{\sqrt{-(g_{tt}+2g_{t\phi}\Omega+g_{\phi\phi}\Omega^{2})}}{1-\Omega b},
\end{equation}
where $b\equiv L_{z}^{\gamma}/E^{\gamma}$, which is a constant of motion along the photon trajectory.

Since the local spectrum is not isotropic, it is necessary to calculate the emission angle. The normal of the disk's surface is given by
\begin{equation}
n^{a}=\frac{1}{\sqrt{g^{rr}Z_{,r}^2+g^{\theta\theta}Z_{,\theta}^2}}(0,g^{rr}Z_{,r},g^{\theta\theta}Z_{,\theta},0)|_{Z(r,\theta)},
\end{equation}
where $Z=Z(r,\theta)$ is the ``surface function'' defined as
\begin{align}
Z(r,\theta)= z(\rho) - 2H =r\cos\theta-\frac{3}{\eta}\left(\frac{\dot{M}}{\dot{M}_{\rm Edd}}\right)\left[1-\sqrt{\frac{r_{\rm ISCO}}{\rho}}\right],
\end{align}
and goes to zero on the surface of the disk. The gradient of $Z(r,\theta)$ gives the normal to the surface. The emission angle is thus given by
\begin{equation}
\cos\theta_{\rm e}=\frac{g}{\sqrt{g^{rr}Z_{,r}^2+g^{\theta\theta}Z_{,\theta}^2}}\left[Z_{,r}\dot{r}+Z_{,\theta}\dot{\theta}\right], \label{eq:cose}
\end{equation}
where $r$ and $\theta$ are the coordinates at the emission point in the disk.

We can define the relative redshift factor $g^{*}$ at a given radius of the accretion disk as~\citep{1975ApJ...202..788C}
\begin{equation}
g^{*}=\frac{g-g_{\text{min}}}{g_{\text{max}}-g_{\text{min}}}\in[0,1],
\end{equation}
where $g_{\text{min}}=g_{\text{min}}(r_{\rm e},\iota)$ and $g_{\text{max}}=g_{\text{max}}(r_{\rm e},\iota)$ represent, respectively, the minimum and maximum values of the redshift factor $g$ for the photons emitted at $r_{\rm e}$ and detected on the distant screen with inclination angle $\iota$.

Introducing the \textit{transfer function}, we can rewrite the observed flux as
\begin{equation}
F_{\rm o}(\nu_{\rm o})= \frac{1}{D^2}
\int^{R_{\text{out}}}_{R_{\text{in}}}\int^{1}_{0}\frac{\pi r_{\rm e}g^{2}f(g^{*},r_{\rm e},\iota)}{\sqrt{g^{*}(1-g^{*})}}I_{\rm e}(r_{\rm e},\theta_{\rm e})dg^{*}dr_{\rm e}, \label{eq:Ione}
\end{equation}
where $R_{\text{out}}$ and $R_{\text{in}}$ are, respectively, the outer and inner radii of the disk. Let us note that we performed a coordinate transformation from $(\alpha,\beta)$ to $(r_e, g^{\ast})$, which means that now we carry out the integration over the accretion disk. $f(g^{*},r_{\rm e},\iota)$ is the transfer function, which is given by
\begin{equation}
f(g^{*},r_{\rm e},\iota)=\frac{1}{\pi r_{\rm e}}g\sqrt{g^{*}(1-g^{*})}\left|\frac{\partial(\alpha,\beta)}{\partial(g^{*},r_{\rm e})}\right|,
\end{equation}
where $\left|\partial(\alpha,\beta)/\partial(g^{*},r_{\rm e})\right|$ is the Jacobian.

\begin{figure*}[t]
\begin{center}
\vspace{0.3cm}
\includegraphics[width=1.05\textwidth,trim={1.0cm 0cm 0cm 0.5cm},clip]{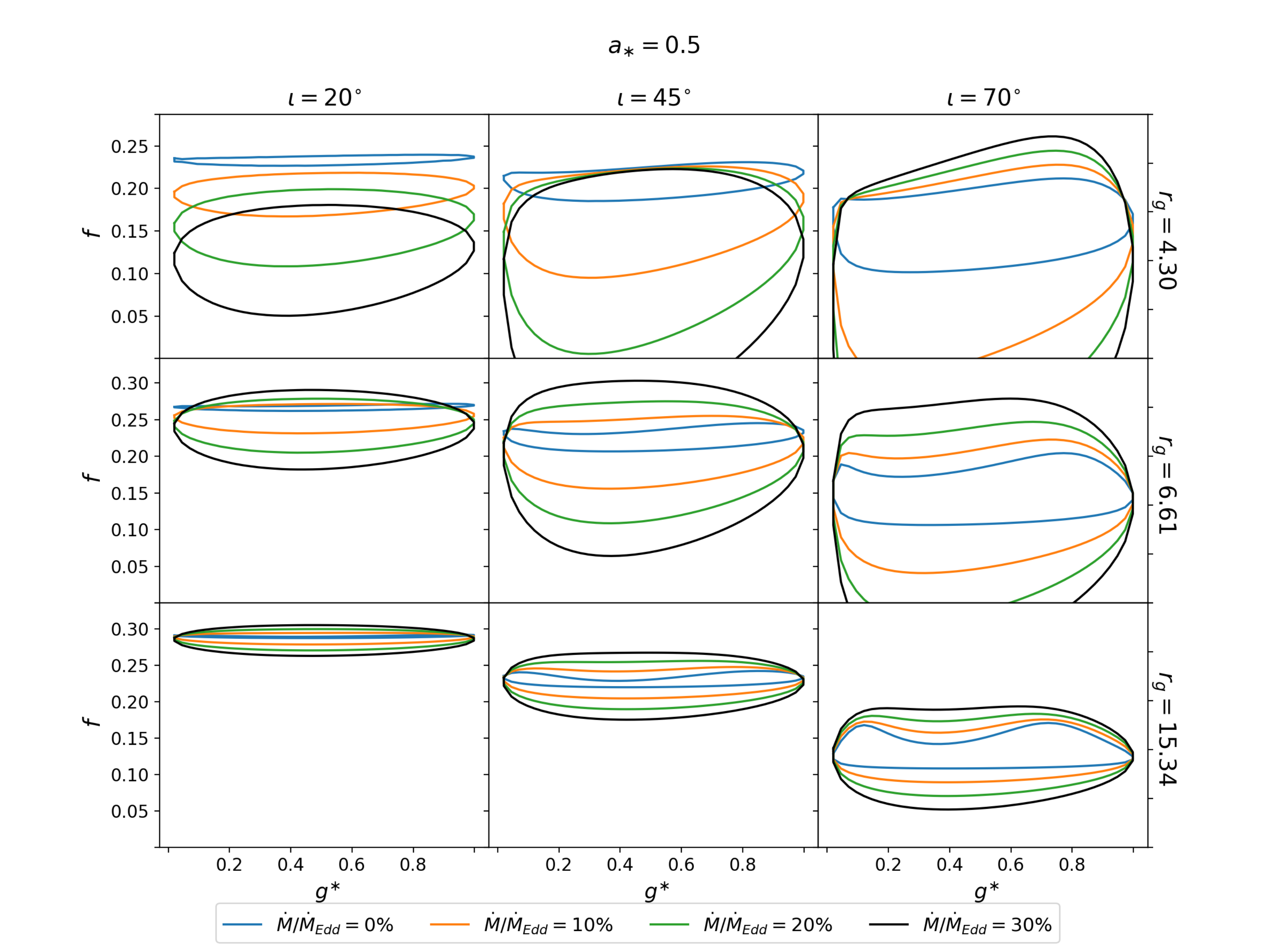}
\end{center}
\caption{Examples of transfer functions at three different radii in Kerr spacetime with spin parameter $a_* = 0.5$ and three different viewing angles ($\iota = 20^\circ$, $45^\circ$, and $70^\circ$, respectively left, central, and right panels). The transfer function of an infinitesimally thin disk ($\dot{M}/\dot{M}_{\rm Edd} = 0$, blue curves) is compared with those of disks of black holes accreting at 10\% (orange curves), 20\% (green curves), and 30\% (black curves) of the Eddington limit. \label{f-tf1}}
\end{figure*}

\begin{figure*}[t]
\begin{center}
\vspace{0.3cm}
\includegraphics[width=1.05\textwidth,trim={1.0cm 0cm 0cm 0.5cm},clip]{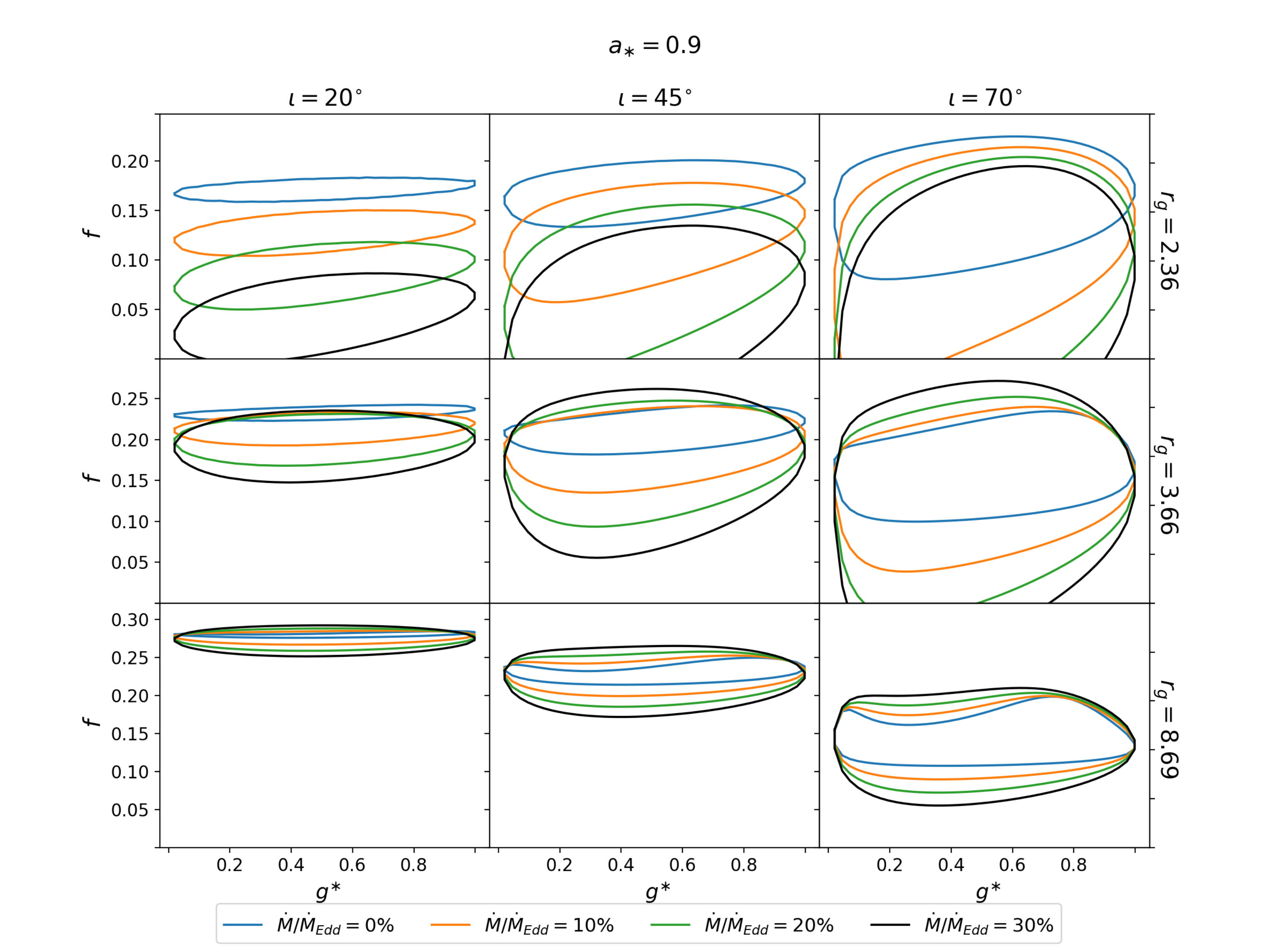}
\end{center}
\caption{As in Fig.~\ref{f-tf1} in Kerr spacetime with spin parameter $a_* = 0.9$. \label{f-tf2}}
\end{figure*}

\begin{figure*}[t]
\begin{center}
\vspace{0.3cm}
\includegraphics[width=1.05\textwidth,trim={1.0cm 0cm 0cm 0.5cm},clip]{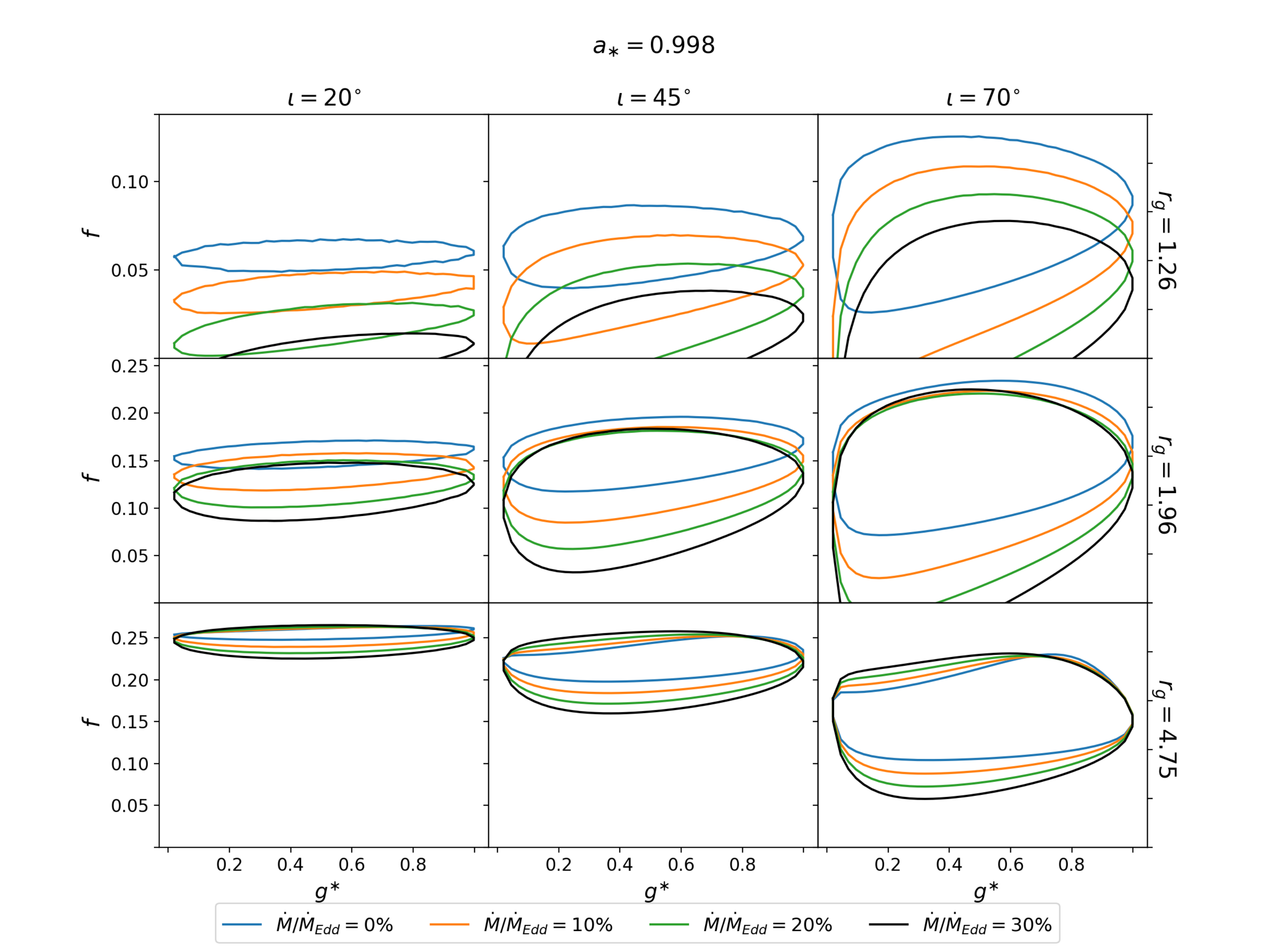}
\end{center}
\caption{As in Fig.~\ref{f-tf1} in Kerr spacetime with spin parameter $a_* = 0.998$. \label{f-tf3}}
\end{figure*}

As noted in~\citet{2018ApJ...855..120T}, the inner part of the accretion disk will be obscured as $\dot{M}/\dot{M}_{\rm Edd}$ increases. For any unobscured part of the disk, for given values of $r_{\rm e}$ and $\iota$, the transfer function is a closed curve parametrized by $g^{\ast}$, except in the special cases $\iota=0$ and $\pi/2$. There is only one point in the disk for which $g^{*}=0$ and one point for which $g^{*}=1$. There are two curves that connect these two points, so there are two branches of the transfer function, say $f^{(1)}(g^{*},r_{\rm e},\iota)$ and $f^{(2)}(g^{*},r_{\rm e},\iota)$. This allows us to rewrite Eq.~(\ref{eq:Ione}) as
\begin{align}\label{flux2}
F_{\rm o}(\nu_{\rm o})&=\frac{1}{D^2} \int^{R_{\text{out}}}_{R_{\text{in}}}\int^{1}_{0}\frac{\pi r_{\rm e}g^{2}f^{(1)}(g^{*},r_{\rm e},\iota)}{\sqrt{g^{*}(1-g^{*})}}I_{\rm e}(r_{\rm e},\theta^{(1)}_{\rm e})dg^{*}dr_{\rm e}
\nonumber \\
&+ \frac{1}{D^2} \int^{R_{\text{out}}}_{R_{\text{in}}}\int^{1}_{0}\frac{\pi r_{\rm e}g^{2}f^{(2)}(g^{*},r_{\rm e},\iota)}{\sqrt{g^{*}(1-g^{*})}}I_{\rm e}(r_{\rm e},\theta^{(2)}_{\rm e})dg^{*}dr_{\rm e},
\end{align}
where $\theta^{(1)}_{\rm e}$ and $\theta^{(2)}_{\rm e}$ present the emission angles with relative redshift factor $g^{*}$ in the branches 1 and 2, respectively. For values of $r_{\rm e}$ for which the disk is obscured, some portion of $f^{(1)}(g^{*},r_{\rm e},\iota)$ and/or $f^{(2)}(g^{*},r_{\rm e},\iota)$ corresponding to the obscured parts of the disk will be equal to zero (see Figs.~\ref{f-tf1}, \ref{f-tf2}, and \ref{f-tf3}), which means there is no radiation contributing from this part into the total reflection spectrum. In such cases, the integration in Eq.~(\ref{flux2}) is performed only on non-zero values of the transfer function.

\subsection{Numerical Method}\label{sec:num}

Here we describe our method for calculating the transfer function and, following the methodology of \textsc{relxill} and \textsc{relxill\_nk}, we create a FITS (Flexible Image Transport System) file that contains the relevant spacetime information. The structure of the FITS file is similar to that used in \textsc{relxill\_nk} for a infinitesimally thin accretion disk and is fully described in~\cite{2019ApJ...878...91A}. There are three physical parameters describing the black hole spacetime in the table, namely, the dimensionless black hole spin parameter $a_*$, the deformation parameter ($\alpha_{13}$ for the metric considered in this paper), and the inclination angle $\iota$,  structured in a 30 by 30 by 22 grid, respectively. The grid for the black hole spin becomes denser as the black hole spin increases, since for high values of $a_*$ the ISCO changes faster. The values of the deformation parameters of the Johannsen metric in the grid are first evenly distributed in the range $[-5,5]$. However, for negative values of $\alpha_{13}$ we may have spacetimes with pathological properties; see Appendix~\ref{app:metric} and the constraint on $\alpha_{13}$ in Eq.~(\ref{eq-constraints}). The lower bound of $\alpha_{13}$ is thus set to the larger value between $-5$ and the bound in Eq.~(\ref{eq-constraints}). The grid point along the inclination angles are evenly distributed in $0<\cos\iota<1$. For each combination of $a_*$, $\alpha_{13}$, and $\iota$, we discretize the accretion disk into 100 emission radii $r_{\rm e}$. For every emission radius, the transfer functions, $f$, and emission angles, $\theta_{\rm e}$, are tabulated at 40 equally spaced values of $g^{\ast}$ on each branch of the transfer function.

As in ~\cite{2019ApJ...878...91A}, a general relativistic ray-tracing code is used to compute the necessary parameters for the FITS file, namely the redshift factor, emission angle, and the Jacobian. The ray-tracing code calculates the trajectories of photons in the Johannsen metric from the black hole accretion disk to a distant observer. The code follows the method described in~\cite{2012ApJ...745....1P} and is a modified version of the one used in~\cite{2019ApJ...878...91A, 2018CQGra..35w5002A, 2019CQGra..36e5007G}. Since it is a stationary and axisymmetric spacetime, the Johannsen metric has a conserved energy $E$ and a conserved angular momentum $L_{z}$. Their relation to the 4-momentum of a test particle, $p_{t}=-E$ and $p_{\phi}=L_{z}$, leads to two first-order differential equations shown in Eqs.~(\ref{eq:dott}) and~(\ref{eq:dotphi}). Rewriting these two equations in terms of the impact parameter $b\equiv L_{z}/E$ and the normalized affine parameter $\lambda'\equiv E\lambda$, 
we obtain
\begin{align}
\frac{dt}{d\lambda'} =& -\frac{bg_{t\phi} + g_{\phi\phi}}{g_{tt}g_{\phi\phi}-g_{t\phi}^{2}}, \label{eq:dt}
\\
\frac{d\phi}{d\lambda'} =& \frac{g_{t\phi}+ b g_{tt}}{g_{tt}g_{\phi\phi}-g_{t\phi}^{2}}. \label{eq:dp}
\end{align}

The second-order geodesic equations for a generic axisymmetric metric describe the evolution of the $r$- and $\theta$-components of the photon's position as
\begin{widetext}
	\begin{align}
	\frac{d^{2}r}{d\lambda'^{2}}=&-\Gamma^{r}_{tt}\left(\frac{dt}{d\lambda'}\right)^{2}-\Gamma^{r}_{rr}\left(\frac{dr}{d\lambda'}\right)^{2}-\Gamma^{r}_{\theta\theta}\left(\frac{d\theta}{d\lambda'}\right)^{2}-\Gamma^{r}_{\phi\phi}\left(\frac{d\phi}{d\lambda'}\right)^{2}-2\Gamma^{r}_{t\phi}\left(\frac{dt}{d\lambda'}\right)\left(\frac{d\phi}{d\lambda'}\right)-2\Gamma^{r}_{r\theta}\left(\frac{dr}{d\lambda'}\right)\left(\frac{d\theta}{d\lambda'}\right), \label{eq:d2r}
	\\
	\frac{d^{2}\theta}{d\lambda'^{2}}=&-\Gamma^{\theta}_{tt}\left(\frac{dt}{d\lambda'}\right)^{2}-\Gamma^{\theta}_{rr}\left(\frac{dr}{d\lambda'}\right)^{2}-\Gamma^{\theta}_{\theta\theta}\left(\frac{d\theta}{d\lambda'}\right)^{2}-\Gamma^{\theta}_{\phi\phi}\left(\frac{d\phi}{d\lambda'}\right)^{2}-2\Gamma^{\theta}_{t\phi}\left(\frac{dt}{d\lambda'}\right)\left(\frac{d\phi}{d\lambda'}\right)-2\Gamma^{\theta}_{r\theta}\left(\frac{dr}{d\lambda'}\right)\left(\frac{d\theta}{d\lambda'}\right), \label{eq:d2th}
	\end{align}
\end{widetext}
where $\Gamma^{a}_{bc}$ indicate the Christoffel symbols of the metric.

A coordinate system and reference frame are chosen in such a way that the black hole is located at the origin and the black hole spin angular momentum is along the $z$-axis. We set $M=1$ in what follows and in the code, since the reflection spectrum does not directly depend on the black hole mass $M$. The observer screen is located far from the black hole at a distance of $D=10^{8}$, with azimuthal angle $\theta=\iota$ and polar angle $\phi=0$. The screen uses the polar coordinates $r_{\text{scr}}$ and $\phi_{\text{scr}}$, and their relation to the celestial coordinates of Eq.~(\ref{eq:celcoords}) are $\alpha=r_{\text{scr}}\cos\phi_{\text{scr}}$ and $\beta=r_{\text{scr}}\sin\phi_{\text{scr}}$.

The code solves the system of equations -- Eqs.~(\ref{eq:dt})-(\ref{eq:d2th}) -- backwards in time. Each photon has an initial position on the screen and an initial 4-momentum perpendicular to the screen. In the Boyer-Lindquist coordinates of the black hole spacetime, the initial position and 4-momentum of the photon are given by
\begin{align}
r_{\rm i}=&\left(\alpha^{2}+\beta^{2}+D^{2}\right)^{1/2},
\\
\theta_{\rm i}=&\arccos\left(\frac{D\cos\iota+\beta\sin\iota}{r_{\rm i}}\right),
\\
\phi_{\rm i}=&\arctan\left(\frac{\alpha}{D\sin\iota-\beta\cos\iota}\right),
\end{align}
and
\begin{align}
\left(\frac{dr}{d\lambda'}\right)_{\rm i}=& \frac{D}{r_{\rm i}},
\\
\left(\frac{d\theta}{d\lambda'}\right)_{\rm i}=&\frac{- \cos\iota + \frac{D}{r_{\rm i}^{2}}\left(D\cos\iota+\beta\sin\iota\right)}{\sqrt{r_{\rm i}^{2}-\left(D\cos\iota+\beta\sin\iota\right)^{2}}},
\\
\left(\frac{d\phi}{d\lambda'}\right)_{\rm i}=&\frac{- \alpha\sin\iota}{\alpha^{2}+\left(D\sin\iota-\beta\cos\iota\right)^{2}},
\\
\left(\frac{dt}{d\lambda'}\right)_{\rm i}=&
-\frac{g_{t\phi}}{g_{tt}}\left(\frac{d\phi}{d\lambda'}\right)_{\rm i}+\sqrt{\frac{g_{t\phi}^{2}}{g_{tt}^{2}}\left(\frac{d\phi}{d\lambda'}\right)_{\rm i}^{2}-\left[\frac{g_{rr}}{g_{tt}}\left(\frac{dr}{d\lambda'}\right)_{\rm i}^{2}
+\frac{g_{\theta\theta}}{g_{tt}}\left(\frac{d\theta}{d\lambda'}\right)_{\rm i}^{2}+\frac{g_{\phi\phi}}{g_{tt}}\left(\frac{d\phi}{d\lambda'}\right)_{\rm i}^{2}\right]}  \, .
\end{align}
The expression for $\left(dt/d\lambda'\right)_{\rm i}$ is obtained by requiring that the norm of the photon 4-momentum vanishes. The impact parameter $b$ is a conserved quantity used in Eqs.~(\ref{eq:dt}) and~(\ref{eq:dp}), calculated from the initial conditions.

Our algorithm adaptively searches for the photons that hit the surface of the accretion disk,~i.e.~the $z(r_{\rm e})$ surface, at the 100 disk emission radii $r_{\rm e}$ to within a precision of $\sim10^{-7}$ varying $r_{\text{scr}}$ and $\phi_{\text{scr}}$. For each emission radius we first shoot 10 photons, from which we register preliminary $g_{\text{min}}$ and $g_{\text{max}}$. Then the actual redshift extremas are found by shifting $\phi_{\text{scr}}$ from these preliminary extremas with an adaptive step-size. Afterwards we search for 80 different photons, 40 in each branch of the transfer function, that correspond to equally distributed values of $g^{\ast}\in[0,1]$. The photons are split into two branches according to
\begin{equation}
\phi_{\text{scr}}^{\text{min}}<\phi_{\text{scr}}<\phi_{\text{scr}}^{\text{max}} \quad
\text{and} \quad
\phi_{\text{scr}}^{\text{min}}>\phi_{\text{scr}}>\phi_{\text{scr}}^{\text{max}}
\end{equation}
where $\phi_{\text{scr}}^{\text{min}}$ and $\phi_{\text{scr}}^{\text{max}}$ are the photons with $g_{\text{min}}$ and $g_{\text{max}}$, respectively.
When searching for photons at a disk emission radius $r_{\rm e}$, we divide photons into real and imaginary ones. Imaginary photons are those that cross the disk several times before landing on the target ring of the disk. Similarly, a real photon does not cross the disk before landing on the target radius. This separation helps to distinguish photons originating from obscured and unobscured parts of the disk. Therefore, imaginary photons originate from the obscured part of the disk that we cannot see. 

For each of these photons, we calculate the redshift factor $g$, Eq.~(\ref{eq:redshift}), emission angle $\theta_{\rm e}$, Eq.~(\ref{eq:cose}), and Jacobian $|\partial(\alpha,\beta)/\partial(g^{*},r_{\rm e})|$. The latter is calculated by using
\begin{equation}
\left|\frac{\partial(\alpha,\beta)}{\partial(g^{*},r_{\rm e})}\right|=(-1)^{p} \left(g_{\text{max}}-g_{\text{min}}\right)  \left[\frac{\partial\alpha}{\partial g}\frac{\partial\beta}{\partial r_{\rm e}}-\frac{\partial\alpha}{\partial r_{\rm e}}\frac{\partial\beta}{\partial g}\right], \label{eq:jac}
\end{equation}
where $p$ is the number of the branch (1 for branch 1 and 2 for branch 2). The introduction of imaginary photons avoids the zero values of the Jacobian, hence the zero values of the transfer function. We attribute a negative sign to them because the negative transfer function allows us to leave unchanged the standard interpolation scheme used in \textsc{relxill\_nk}. We set negative values of the transfer function to zero only after the final interpolation in the model. The third term on the right-hand side is computed using an adaptive algorithm that, when solving the geodesic equations, searches for two photons that have $g\pm\Delta g$ for the given emission radius $r_{\rm e}$, where $g$ is the initial redshift factor of the original and $\Delta g<10^{-6}$. For $r_{\rm e} \pm \Delta r$, the code uses adjacent photons from two neighboring emission rings, thus $\Delta r$ is the distance between these photons. The derivatives are then calculated from the emission radius, redshift factor, and initial coordinates of these four photons in a separate code, as the second term on the right-hand side of Eq.~(\ref{eq:jac}). 

Finally, we use a separate script to process all photons and create the FITS file for a specific value of $\dot{M}/\dot{M}_{\rm Edd}$. First, the script calculates the Jacobian and then generates a FITS file containing the values of 100 emission radii $r_{\rm e}$, corresponding minimum and maximum redshift values ($g_{\text{min}}$ and $g_{\text{max}}$), transfer functions, and emission angles $\theta_{\rm e}$, for the full combination of dimensionless spin $a_{*}$, deformation parameter $\alpha_{13}$, and inclination angle $\iota$.

\subsection{Impact of disk thickness on the iron lines}\label{sec:imp-iron-lines}

\begin{figure*}[t]
\begin{center}
\vspace{0.3cm}
\includegraphics[width=1.05\textwidth,trim={1.0cm 0cm 0cm 0.5cm},clip]{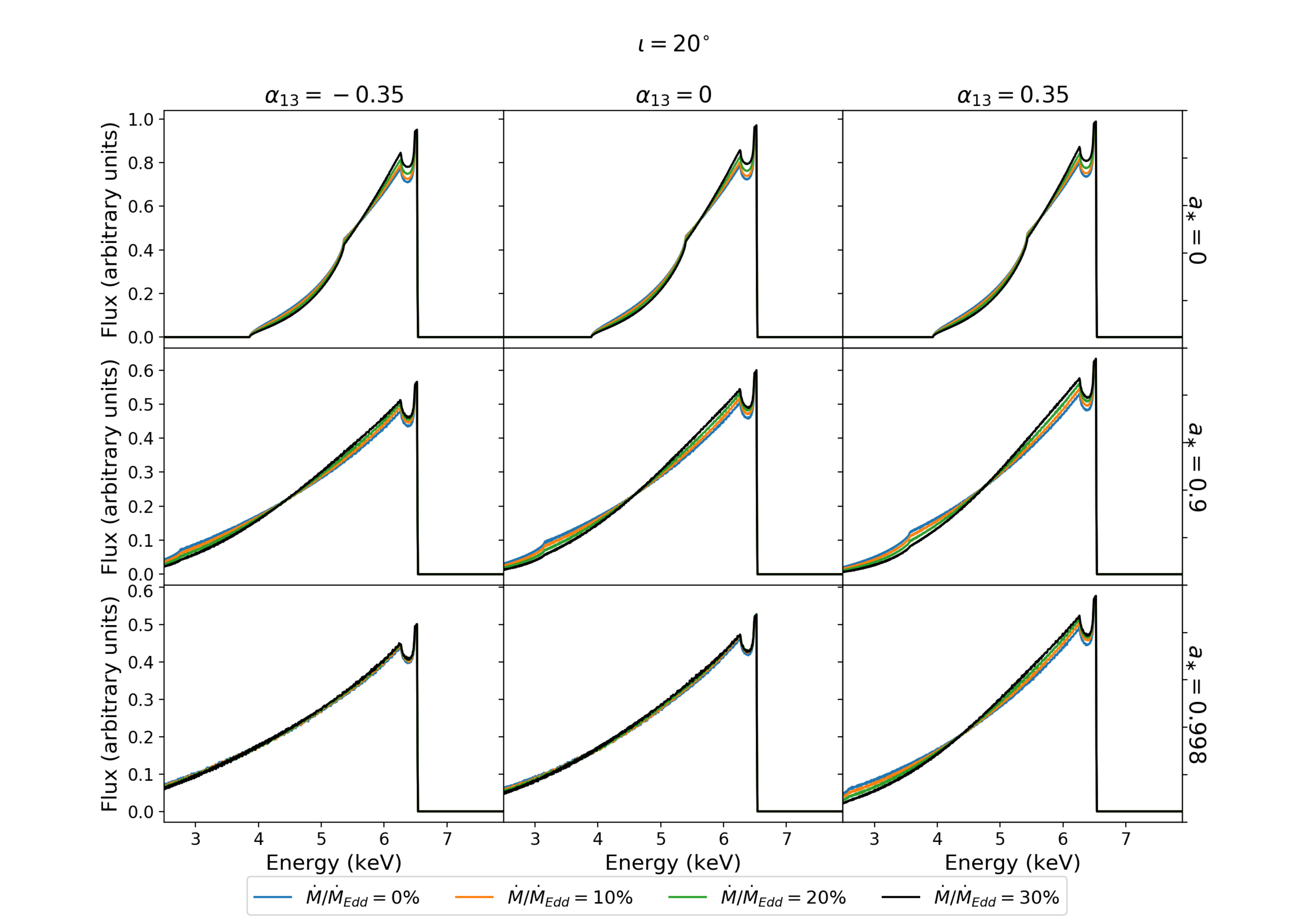}
\end{center}
\caption{Examples of iron line profiles in the Johannsen metric for a spin parameter $a_* = 0$, 0.9, and 0.998, a deformation parameter $\alpha_{13} = -0.35$, 0, 0.35, and an inclination angle $\iota = 20^\circ$. The iron line profile for an infinitesimally thin disk ($\dot{M}/\dot{M}_{\rm Edd} = 0$, blue profiles) is compared with those for black holes accreting at 10\% (orange profiles), 20\% (green profiles), and 30\% (black profiles) of the Eddington limit. \label{f-iron1}}
\vspace{0.4cm}
\end{figure*}

\begin{figure*}[t]
\begin{center}
\vspace{0.3cm}
\includegraphics[width=1.05\textwidth,trim={1.0cm 0cm 0cm 0.5cm},clip]{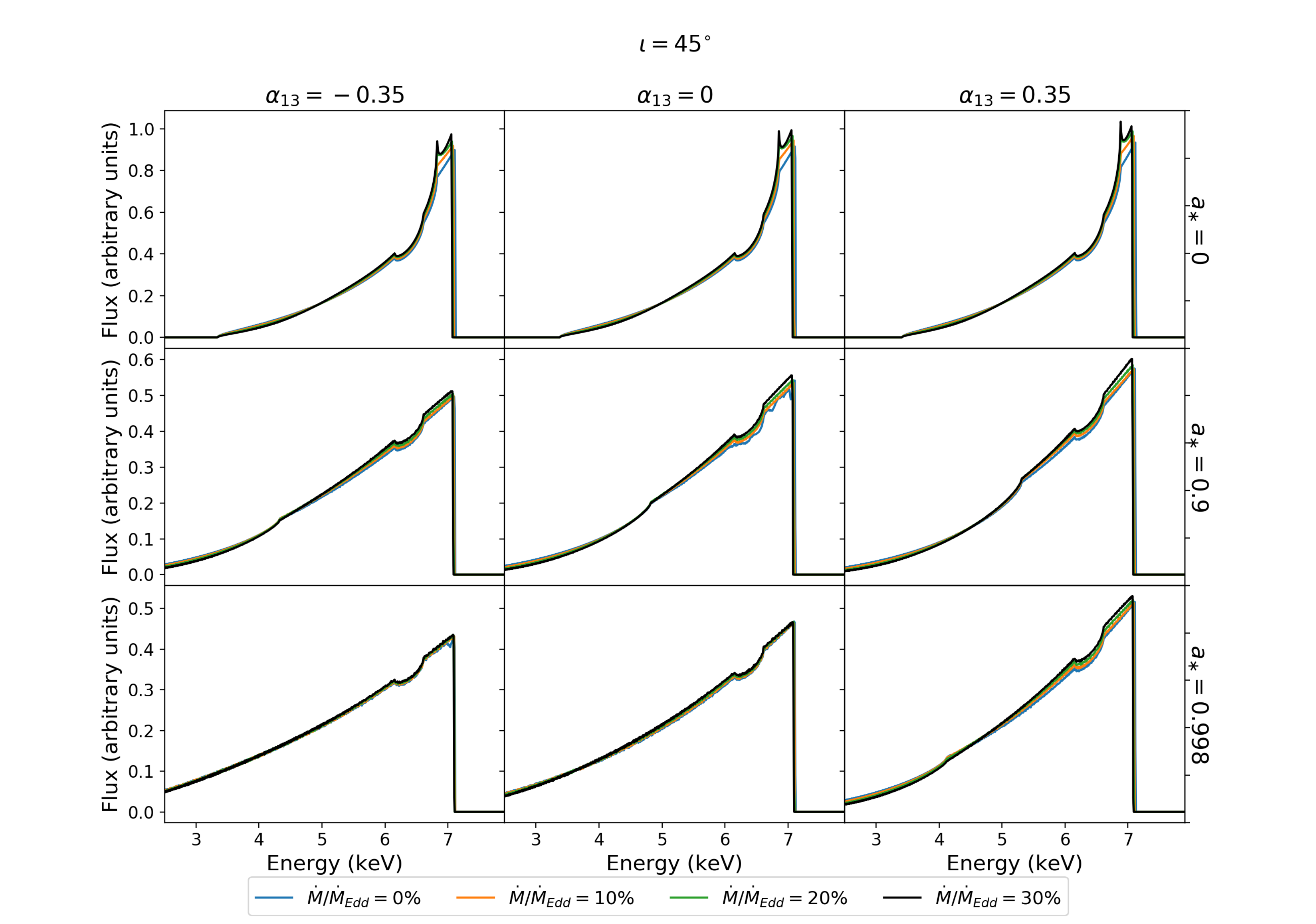}
\end{center}
\caption{As in Fig.~\ref{f-iron1} for a viewing angle $\iota = 45^\circ$. \label{f-iron2}}
\vspace{0.4cm}
\end{figure*}

\begin{figure*}[t]
\begin{center}
\vspace{0.3cm}
\includegraphics[width=1.05\textwidth,trim={1.0cm 0cm 0cm 0.5cm},clip]{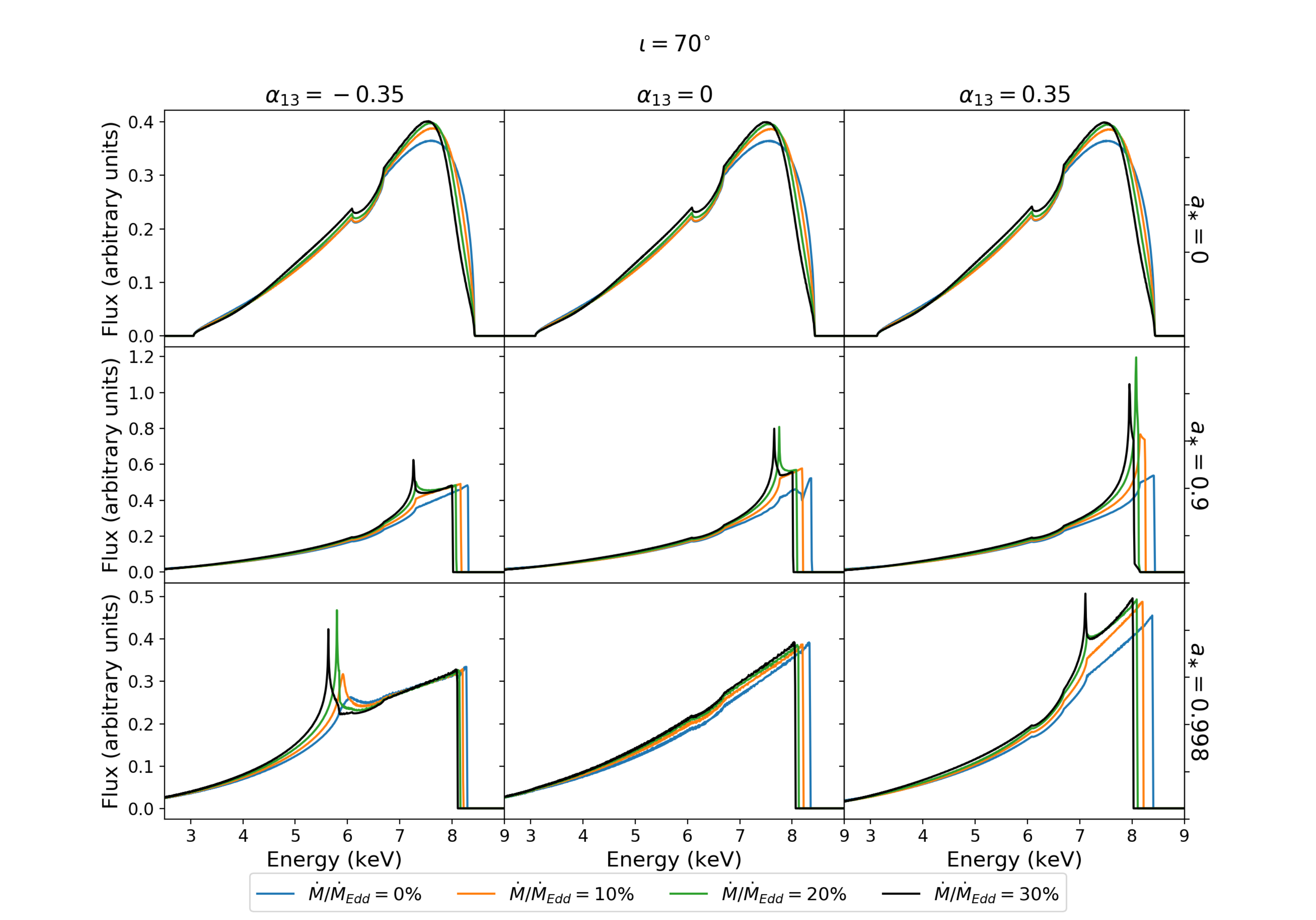}
\end{center}
\caption{As in Fig.~\ref{f-iron1} for a viewing angle $\iota = 70^\circ$. \label{f-iron3}}
\vspace{0.4cm}
\end{figure*}

Figs.~\ref{f-iron1}, \ref{f-iron2}, and \ref{f-iron3} show iron line profiles for different values of $a_*$, $\alpha_{13}$, $\iota$, and $\dot{M}/\dot{M}_{\rm Edd}$. All calculations assume that the energy of the line in the rest-frame of the gas is $E = 6.4$~keV and that the intensity profile of the disk is described by a power-law with emissivity index $q = 3$, i.e. $I_{\rm e} \propto \rho^{-3}$. Here the intensity profile is described using the coordinate $\rho$, and not $r$, because such a choice guarantees a one-to-one correspondence between every value of $\rho$ and every point of the surface of the disk (modulo the cylindrical symmetry of the system). At a qualitative level, we can say that the impact of the thickness of the disk on the iron line profile is weak for $\iota = 20^\circ$ and $45^\circ$ (Figs.~\ref{f-iron1} and \ref{f-iron2}), while it is a bit larger for $\iota = 70^\circ$ (Fig.~\ref{f-iron3}). For $a_*$ and $\alpha_{13}$ there is not a clear trend: generally speaking, if the ISCO radius is closer to the black hole, the gravitational field is stronger, and we can expect that small variations in the exact emission point has a larger impact on the shape of the iron line profile; however, when the ISCO radius is closer to the black hole, $\eta$ is typically higher, making the disk thinner for the same value of $\dot{M}/\dot{M}_{\rm Edd}$, thus producing the opposite effect with smaller difference with respect to an infinitesimally thin disk.

\begin{figure*}[t]
\begin{center}
\vspace{0.0cm}
\includegraphics[width=0.45\textwidth,trim={0cm 0cm 0.5cm 0cm},clip]{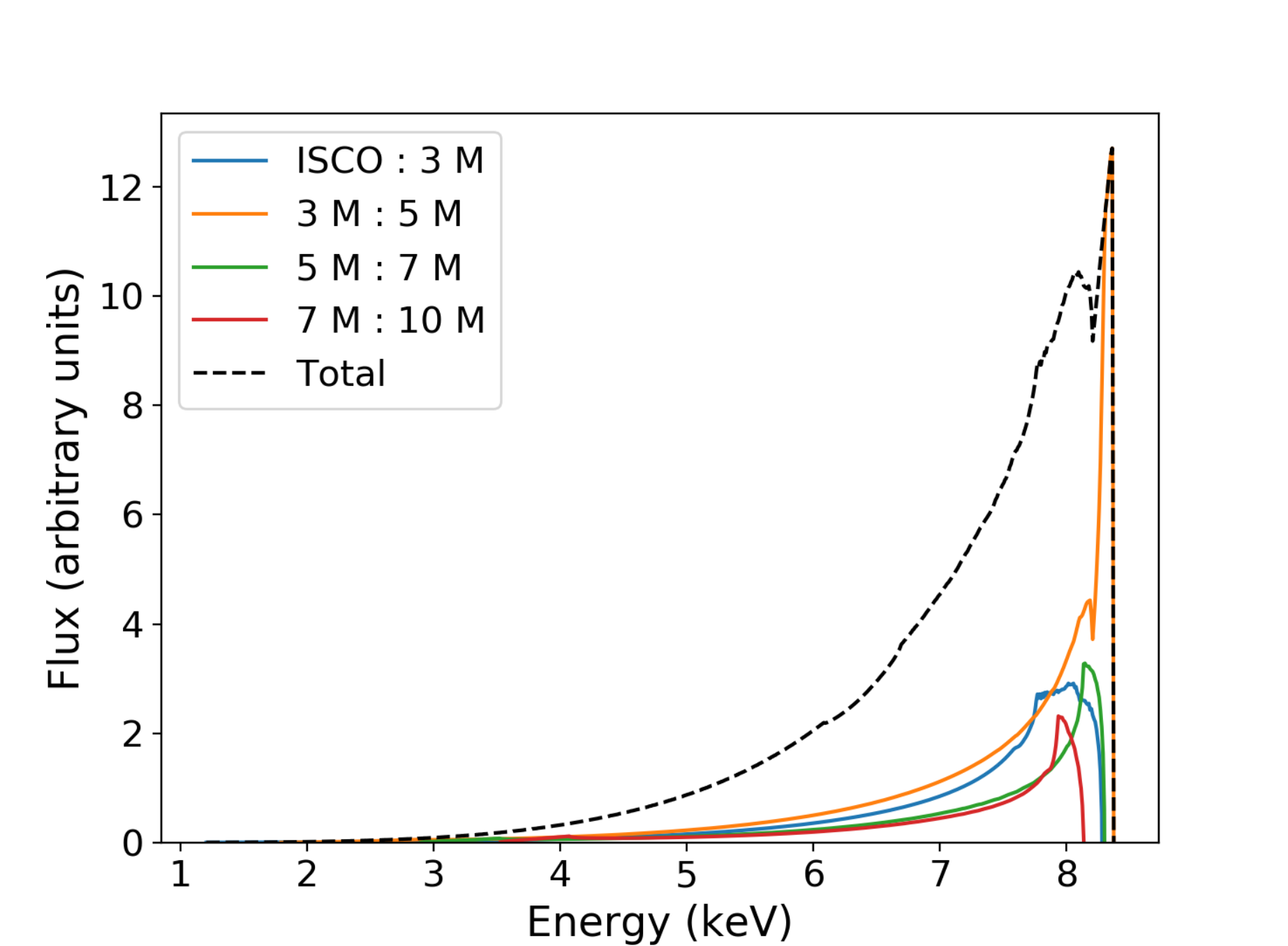}
\includegraphics[width=0.45\textwidth,trim={0cm 0cm 0.5cm 0cm},clip]{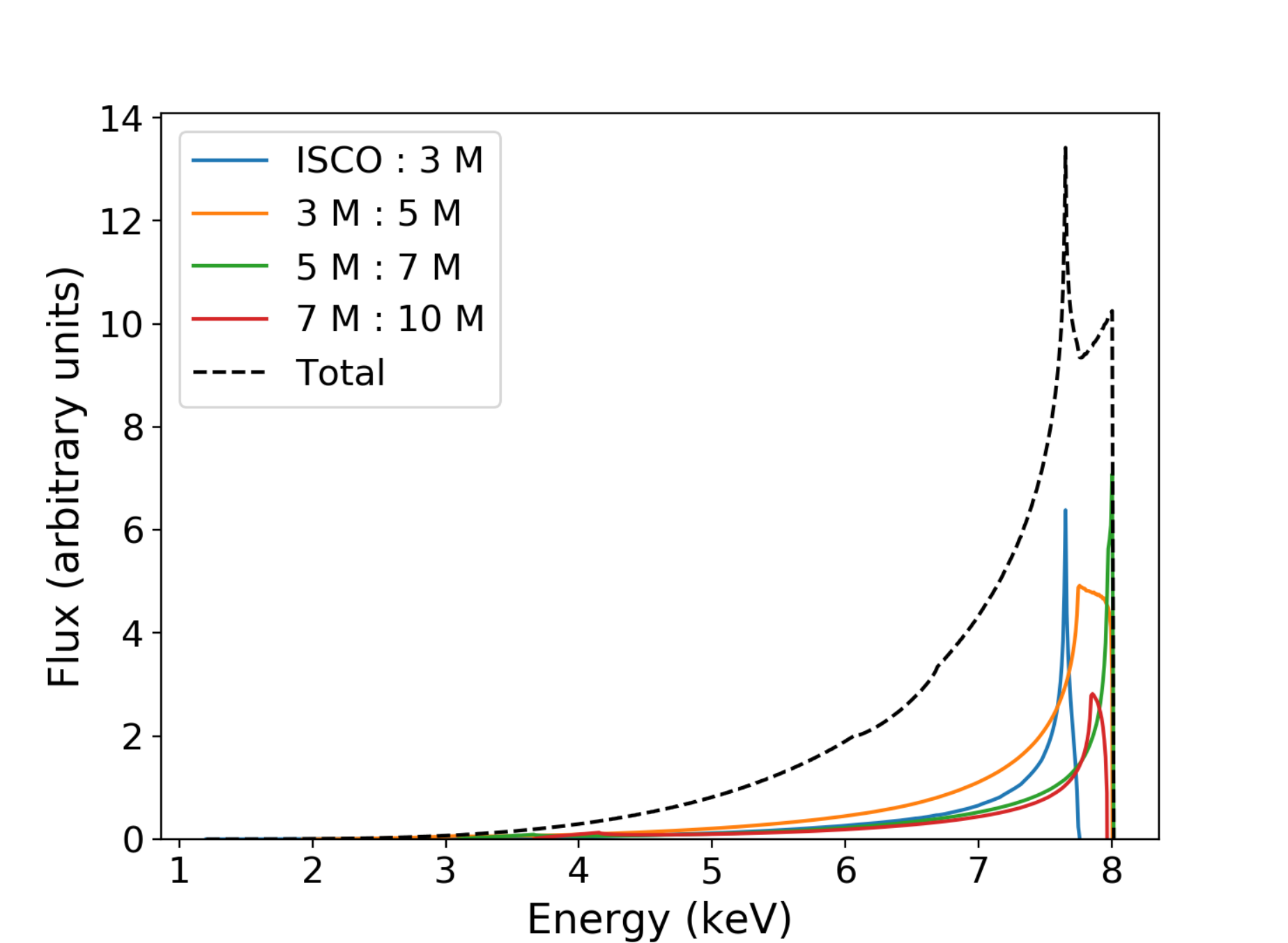}
\end{center}
\caption{Total iron line profile and contributions from a few inner annuli of an infinitesimally thin accretion disk (left panel) and of an accretion disk of finite thickness with $\dot{M}/\dot{M}_{\rm Edd} = 0.3$ (right panel). The annuli are: $r_{\rm ISCO} < \rho < 3$~$M$ (blue curves), 3~$M < \rho < 5$~$M$ (yellow curves), 5~$M < \rho < 7$~$M$ (green curves), and 7~$M < \rho < 10$~$M$ (red curves). The spacetime is described by the Kerr metric with $a_* = 0.9$, the viewing angle is $\iota = 70^\circ$, and the emissivity index of the intensity profile is $q = 3$. \label{f-deco}}
\vspace{0.4cm}
\end{figure*}

\begin{figure*}[t]
\begin{center}
\vspace{0.3cm}
\includegraphics[width=1.05\textwidth,trim={1.0cm 0cm 0cm 0.5cm},clip]{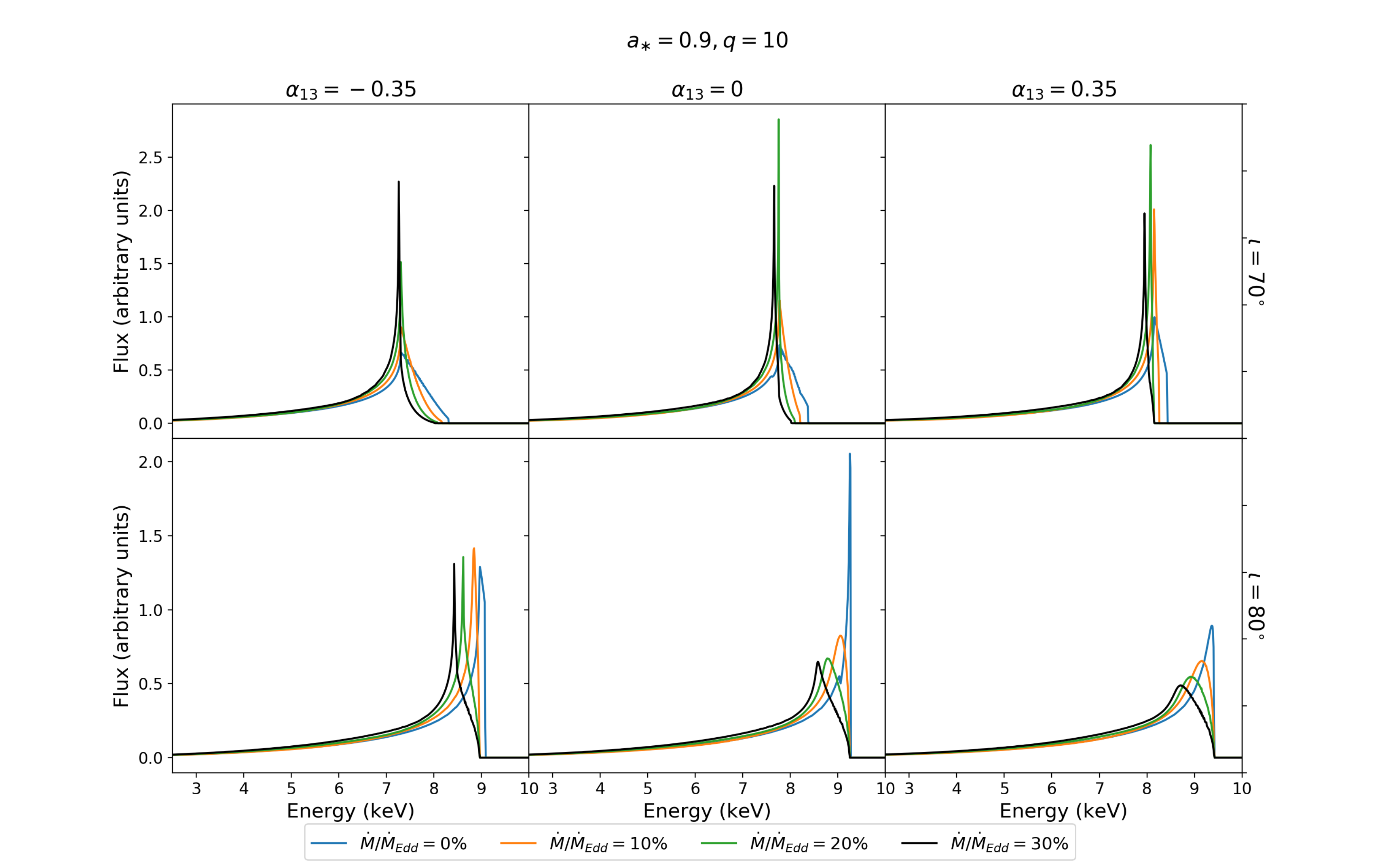}
\end{center}
\caption{Examples of iron line profiles in the Johannsen metric for a spin parameter $a_* = 0.9$, a deformation parameter $\alpha_{13} = -0.35$, 0, 0.35, an inclination angle $\iota = 70^\circ$ and $80^\circ$, and an emissivity profile $q=10$. The iron line profile for an infinitesimally thin disk ($\dot{M}/\dot{M}_{\rm Edd} = 0$, blue profiles) is compared with those for black holes accreting at 10\% (orange profiles), 20\% (green profiles), and 30\% (black profiles) of the Eddington limit. \label{f-iron4}}
\vspace{0.4cm}
\end{figure*}

\begin{figure*}[t]
\begin{center}
\vspace{0.3cm}
\includegraphics[width=1.05\textwidth,trim={1.0cm 0cm 0cm 0.5cm},clip]{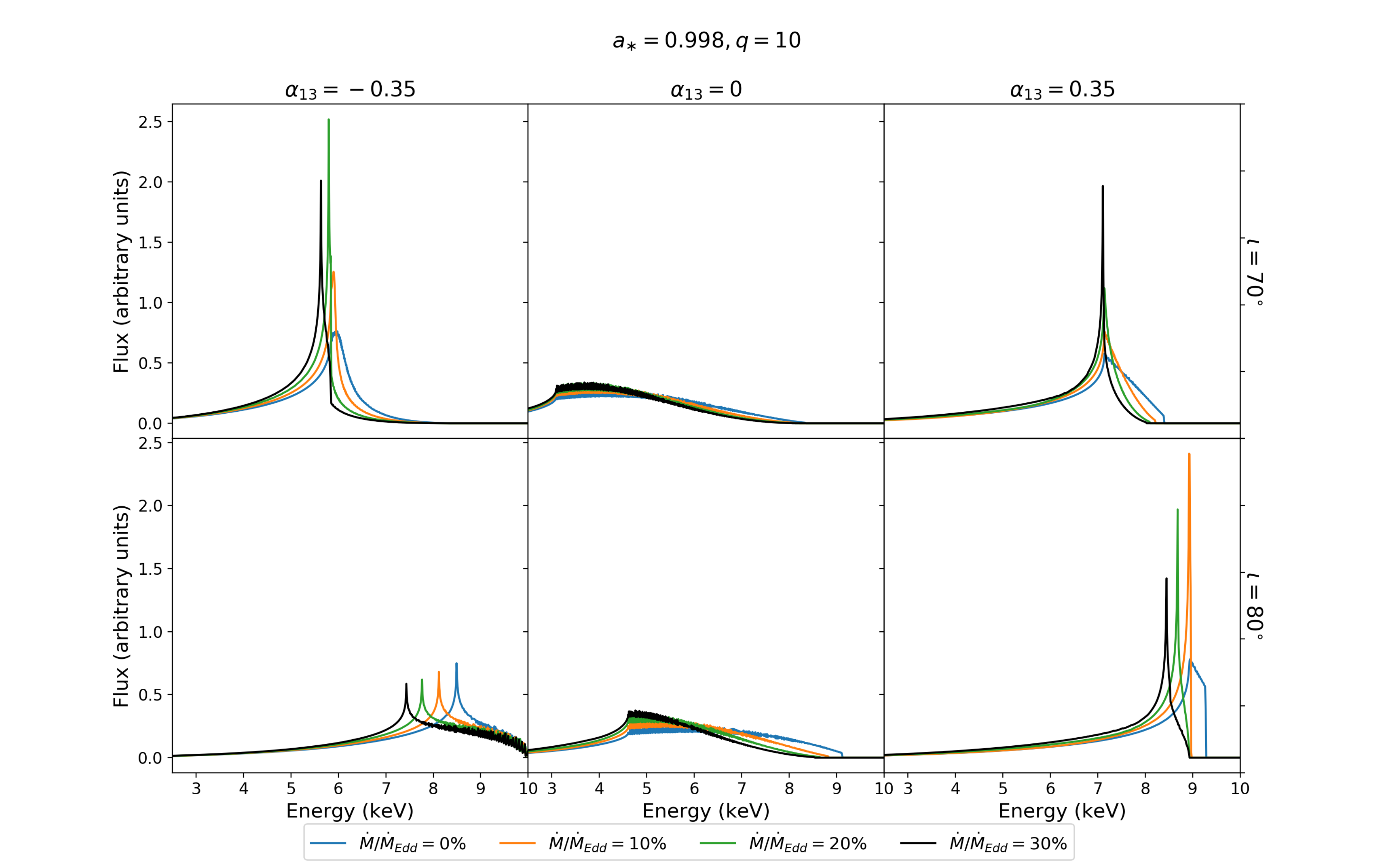}
\end{center}
\caption{As in Fig.~\ref{f-iron4} for a spin parameter $a_* = 0.998$.  \label{f-iron5}}
\vspace{0.4cm}
\end{figure*}

In Fig.~\ref{f-iron3}, we see a peaky feature in some iron line profiles for disks of finite thickness, while such a feature is never present in the case of infinitesimally thin disks. It is not a numerical artifact but the result of a combination of the disk self-shadowing, namely the fact that a disk of finite thickness can obscure a portion of the very inner part of the disk, and relativistic effects. We note that the effect of self-shadowing shows up above some critical value of the viewing angle (which depends on the black hole spin, deformation parameter, mass accretion rate, and emissivity profile) and the effect is more and more important as the viewing angle increases, as both the shelf-shadowing and the Doppler boosting increase. In the Kerr spacetime ($\alpha_{13} = 0$) we see a peaky feature in Fig.~\ref{f-iron3}, where $\iota = 70^\circ$, and for $a_* = 0.9$. When $a_* = 0.998$, the peaky feature disappears because the radiative efficiency $\eta$ is significantly higher and the disk is thus thinner, see Eq.~(\ref{eq:Hdef}) and Fig.~\ref{fig:disk}. \citet{2018ApJ...855..120T} do not find such a peaky feature in their paper because their plots show iron line profiles observed with viewing angles up to $60^\circ$, while Fig.~\ref{f-iron3} is for $70^\circ$. Moreover, the feature depends also on the disk intensity profile: here we use a power law profile while \citet{2018ApJ...855..120T} employ a lamppost emissivity profile.

To illustrate the origin of the peaky feature, in Fig.~\ref{f-deco} we show the iron line profile contributions from a few inner annuli of the accretion disk for $\dot{M}/\dot{M}_{\rm Edd} = 0$ (left panel) and $\dot{M}/\dot{M}_{\rm Edd} = 0.3$ (right panel) in the case $a_* = 0.9$, $\iota = 70^\circ$, and $q = 3$. As we can see from the comparison of the two panels, the spectra of the inner annuli ($r_{\rm ISCO} < \rho < 3$~$M$) is significantly different in the two iron line profiles, while the spectra of outer annuli are quite similar. The spectra of annuli with $r > 10$~$M$ are not shown in Fig.~\ref{f-deco} because the difference is negligible. While the spectrum of the inner annulus of the infinitesimally thin disk has a broad peak, that of the disk with finite thickness has a sharp cut-off at high energy due to the obscuration of the disk. Doppler effect and light bending are also different as the gas keeps the same angular velocity but leaves the equatorial plane of the infinitesimally thin disk; indeed the emission angle $\theta_{\rm e}$ changes and the inner part of the accretion disk behind the black hole becomes more face on as the thickness of the disk increases, affecting the Doppler boosting.

Fast-rotating black holes observed from large viewing angles are the most interesting and suitable sources for testing the Kerr metric, since these two properties maximize the relativistic effects in the reflection spectrum. As we have seen in this section, these turn out to be even the sources with the reflection spectra more affected by the thickness of the disk. It is thus important to further investigate the features of their spectra. In particular, a relevant parameter is the emissivity profile. In Figs.~\ref{f-iron1}-\ref{f-iron3}, we have calculated iron line profiles employing an emissivity index $q=3$. In the analysis of real data, it is common to find much higher emissivity indices, especially for the inner part of the disk (an example is presented in the next section). If we increase the value of $q$, we increase the fraction of radiation emitted from the very inner part of the accretion disk, enhancing the impact of the thickness of the disk on the iron line shape.

Fig.~\ref{f-iron4} shows iron line shapes for a spin parameter $a_*=0.9$, a viewing angle $\iota = 70^\circ$ and $80^\circ$, and an emissivity index $q=10$ for infinitesimally thin disks and disks of finite thickness. Fig.~\ref{f-iron5} is as Fig.~\ref{f-iron4} but for $a_*=0.998$. We note that employing $q=6$ does not change much the line shapes, while there are clear differences for lower values of $q$. While the effects of $a_*$, $\alpha_{13}$, $\iota$, $q$, and $\dot{M}/\dot{M}_{\rm Edd}$ mix together in a quite complicated way, we can see that the main impact of the disk thickness is to move the peak of the iron line shape to lower energies. This is again the effect of the obscuration of the inner part of the accretion disk and of the different Doppler boosting discussed a few paragraphs above and related to the peaky feature of the disk with finite thickness. Enhancing the fraction of radiation emitted from the very inner part of the accretion disk by increasing the value of $q$ leads to increase the effect of the thickness of the disk on the shape of the iron line.


\begin{table*}
\centering
\caption{ \label{t-fit}}
{\renewcommand{\arraystretch}{1.4}
\begin{tabular}{lcc}
\hline\hline
 & Infinitesimally thin disk & Disk with finite thickness \\
\hline
{\sc tbabs} && \\
$N_{\rm H} / 10^{22}$ cm$^{-2}$ & $7.97^{+0.07}_{-0.09}$ & $7.867^{+0.022}_{-0.024}$ \\
\hline
{\sc relxill\_nk} && \\
$q_{\rm in}$ & $10.0_{-0.6}$ & $8.55^{+0.13}_{-1.01}$ \\
$q_{\rm out}$ & $0.00^{+0.21}$ & $0.0^{+1.1}$ \\
$R_{\rm br}$~$[M]$ & $6.03^{+0.18}_{-0.44}$ & $7.26^{+3.62}_{-0.11}$ \\
$i$ [deg] & $73.7^{+1.6}_{-0.6}$ & $79.6^{+3.3}_{-0.5}$ \\
$a_*$ & $0.9897^{+0.0015}_{-0.0009}$ & $0.9950^{\rm (P)}_{-0.0003}$ \\
$\alpha_{13}$ & $-0.09^{+0.10}_{-0.10}$ & $0.00^{+0.05}_{-0.15}$ \\
$\dot{M}/\dot{M}_{\rm Edd}$ & $0^\star$ & $0.2^\star$ \\
$\log\xi$ & $2.77^{+0.03}_{-0.04}$ & $2.699^{+0.011}_{-0.010}$ \\
$A_{\rm Fe}$ & $0.60^{+0.07}_{-0.06}$ & $0.737^{+0.021}_{-0.032}$ \\
$\Gamma$ & $2.199^{+0.015}_{-0.016}$ & $2.2120^{+0.0059}_{-0.0016}$ \\
$E_{\rm cut}$ [keV] & $71.2^{+3.3}_{-1.6}$ & $69.6^{+0.5}_{-1.1}$ \\
$R_{\rm f}$ & $0.48^{+0.09}_{-0.03}$ & $0.461^{+0.006}_{-0.073}$ \\
norm & $0.0429_{-0.0025}^{+0.0004}$ & $0.04626_{-0.0044}^{+0.0005}$ \\
\hline
$\chi^2/\nu$ & $\quad 2314.75/2208 \quad$ & $\quad 2306.52/2208 \quad$ \\
& =1.04835 & =1.04462 \\
\hline\hline
\end{tabular}}
\vspace{0.2cm}
\tablenotetext{0}{Best-fit values from the analysis of the 2007 \textsl{Suzaku} observation of GRS~1915+105 with {\sc relxill\_nk} employing an infinitesimally thin disk (left column) and a disk with finite thickness for $\dot{M}/\dot{M}_{\rm Edd} = 0.2$. The reported uncertainties correspond to the 90\% confidence level for one relevant parameter ($\Delta\chi^2 = 2.71$). $^\star$ indicates that the parameter is frozen in the fit. Note that $q_{\rm in}$ and $q_{\rm out}$ are allowed to vary in the range [0,10] and the best-fits are stuck at the boundary with the exception of $q_{\rm in}$ for the model with $\dot{M}/\dot{M}_{\rm Edd} = 0.2$. The maximum value of the spin parameter allowed by the model is 0.998, and for $\dot{M}/\dot{M}_{\rm Edd} = 0.2$ the 90\% confidence level reaches the boundary.} 
\vspace{0.2cm}
\end{table*}

\section{Impact of the disk thickness on tests of the Kerr black hole hypothesis}\label{sec:grs1915}

With the FITS file for a specific value of $\dot{M}/\dot{M}_{\rm Edd}$, we can test the new model with real data in order to estimate the systematic uncertainties of the model with an infinitesimally thin disk and the impact of the disk thickness on our tests of the Kerr black hole hypothesis. For a preliminary study to present in this paper, we consider the 2007 \textsl{Suzaku} observation of GRS~1915+105, which was analyzed for the first time by~\citet{2009ApJ...706...60B} with a Kerr model and was analyzed by our group in~\citet{2019ApJ...884..147Z} to test the Kerr metric with {\sc relxill\_nk}. These data provide, as of now, one of the most stringent constraints on $\alpha_{13}$ among all the observations and sources analyzed so far with {\sc relxill\_nk}, and they are thus suitable to test the impact of the disk thickness on the measurement of the deformation parameter $\alpha_{13}$. Note that during the 2007 \textsl{Suzaku} observation, the Eddington-scaled accretion luminosity of the source was about 0.2\footnote{\citet{2009ApJ...706...60B} estimate the Eddington-scaled accretion luminosity of the source $\sim$0.3, but they assume the black hole mass $M= 14 \pm 4$~$M_\odot$~\citep{2001A&A...373L..37G} and distance $D \sim 12$~kpc~\citep{2004ARA&A..42..317F}. Here we use the more recent measurements reported in~\citet{2014ApJ...796....2R}: $M = 12.4_{-1.8}^{+2.0}$~$M_\odot$ and $D = 8.6_{-1.6}^{+2.0}$~kpc.}, which is a high value but still in the range expected for a geometrically thin accretion disk with inner edge at the ISCO radius.

The observation, data reduction, and choice of the model was already discussed in~\citet{2019ApJ...884..147Z}. Here we just point out the main properties of this observation and this source. GRS~1915+105 is quite a bright stellar-mass black hole. Previous analyses in Kerr and Johannsen backgrounds suggest that the inclination angle of the disk is high (around $70^\circ$) and the inner edge of the accretion disk is very close to the black hole. A high value of the inclination angle tends to maximize the relativistic effects, in particular the light bending. Moreover, as we have seen at the end of the previous section, it seems that a high inclination angle maximizes the impact of the thickness of the accretion disk, which is indeed what we want to explore here. Concerning the position of the inner edge of the accretion disk, we meet two opposing effects. If the inner edge is closer to the black hole, the signature of the strong gravity effects in the reflection spectrum are larger, and it may be possible that small differences in the location of the emission can have an impact on the reflection spectrum. Note also that in our tests of the Kerr black hole hypothesis we typically prefer to analyze sources with an inner edge of the accretion disk very close to the compact object because this helps to break the parameter degeneracy and constrain the deformation parameter; if the inner edge of the accretion disk is far from the source, simultaneous measurements of the black hole spin and the deformation parameter are difficult or impossible. On the contrary, an inner edge of the accretion disk very close to the black hole is typically accompanied by a high value of the radiative efficiency $\eta$, which makes the disk thinner via Eq.~(\ref{eq:Hdef}). The difference with the infinitesimally thin disk should thus be smaller.

\begin{figure*}[t]
\begin{center}
\includegraphics[width=0.45\textwidth,trim={2.0cm 0cm 3.0cm 17.0cm},clip]{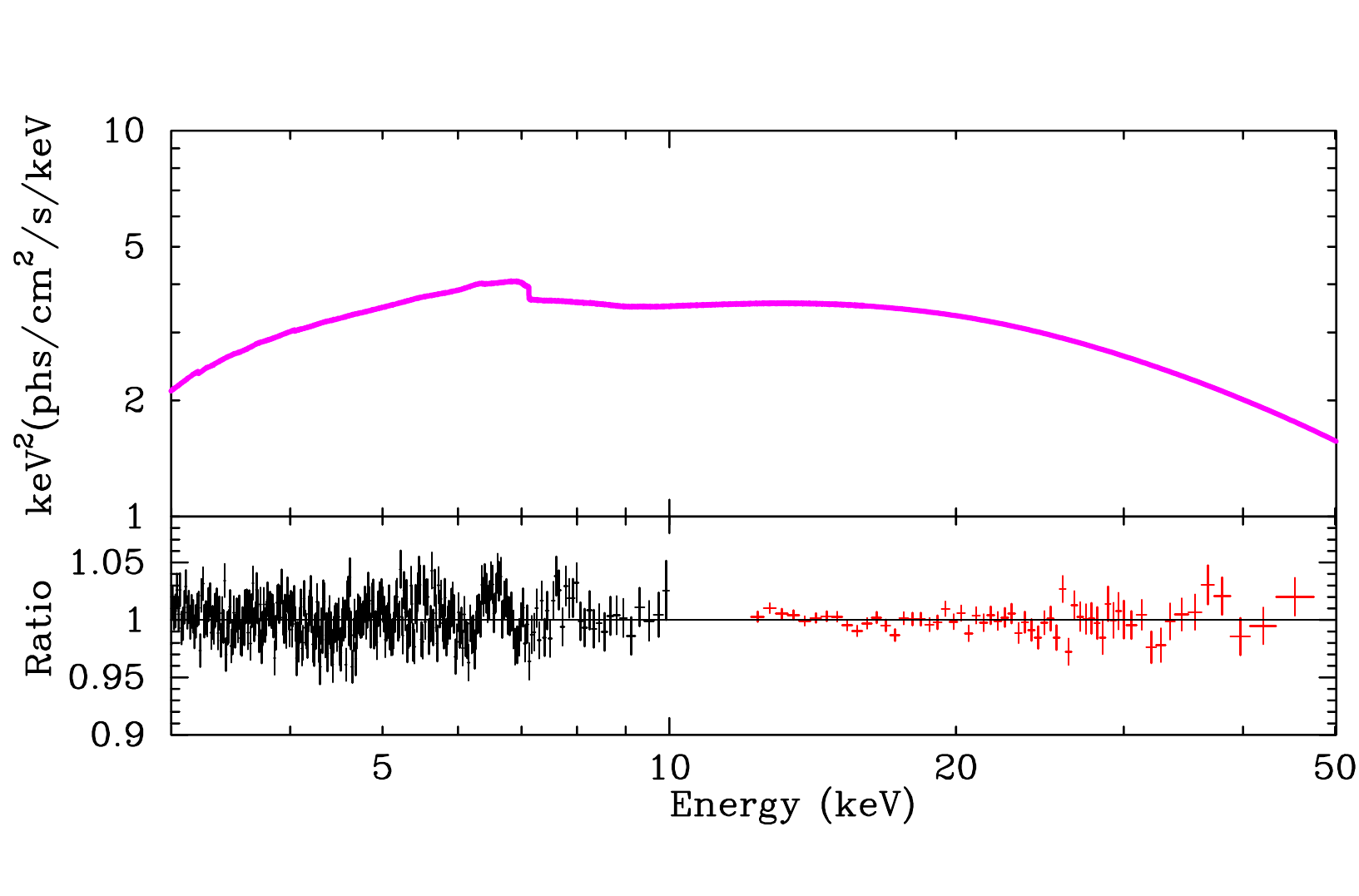}
\hspace{0.5cm}
\includegraphics[width=0.45\textwidth,trim={2.0cm 0cm 3.0cm 17.0cm},clip]{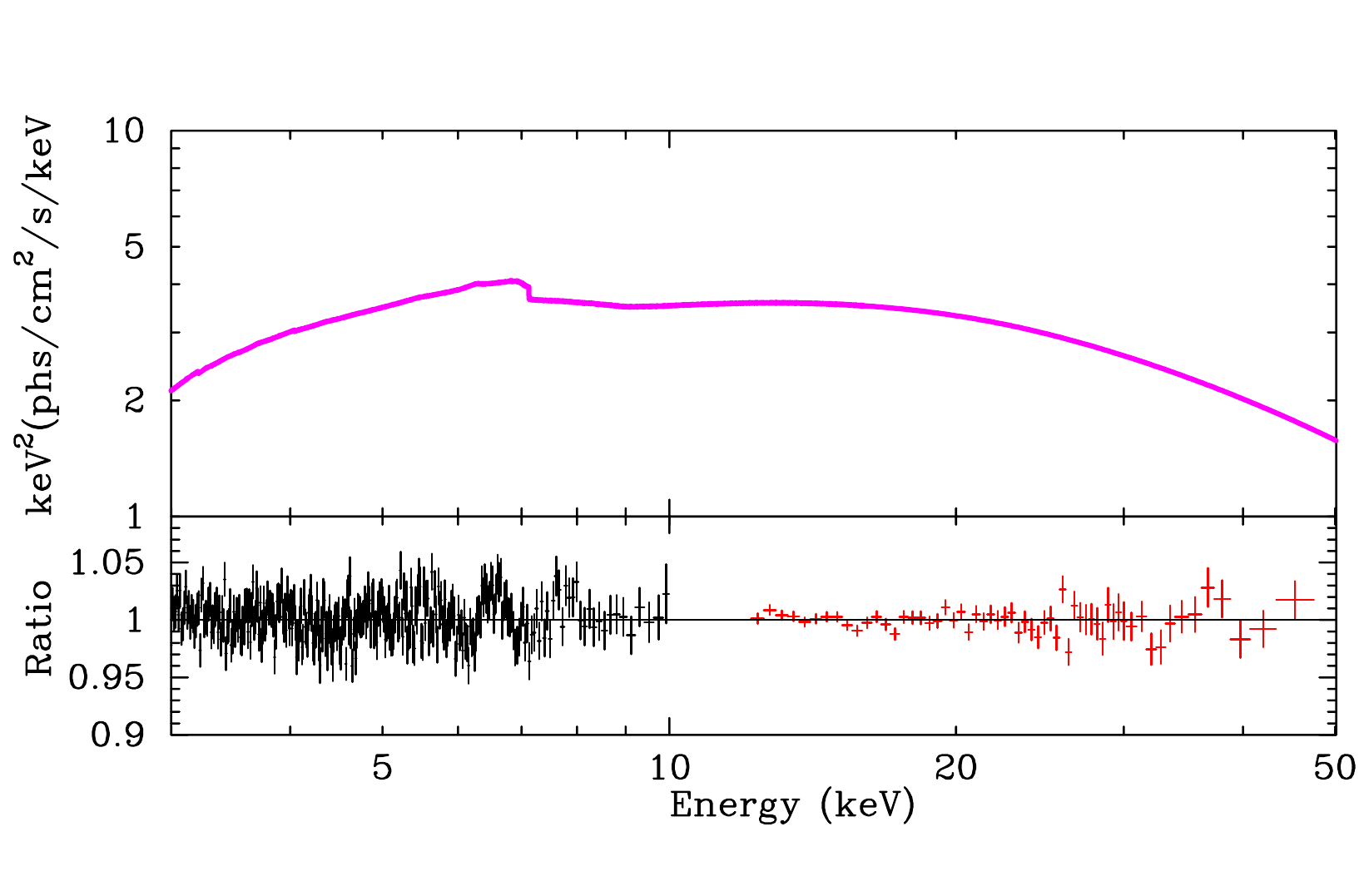}
\end{center}
\vspace{-0.5cm}
\caption{Best-fit models (top quadrants) and data to best-fit model ratios (bottom quadrants) for the models with an infinitesimally thin disk (left panel) and finite disk thickness with $\dot{M}/\dot{M}_{\rm Edd} = 0.2$ (right panel). The black and red crosses are, respectively, for XIS and PIN data. \label{f-ratio}}
\end{figure*}

\begin{figure*}[t]
\begin{center}
\includegraphics[width=0.45\textwidth,trim={1.5cm 2.0cm 1.0cm 0cm},clip]{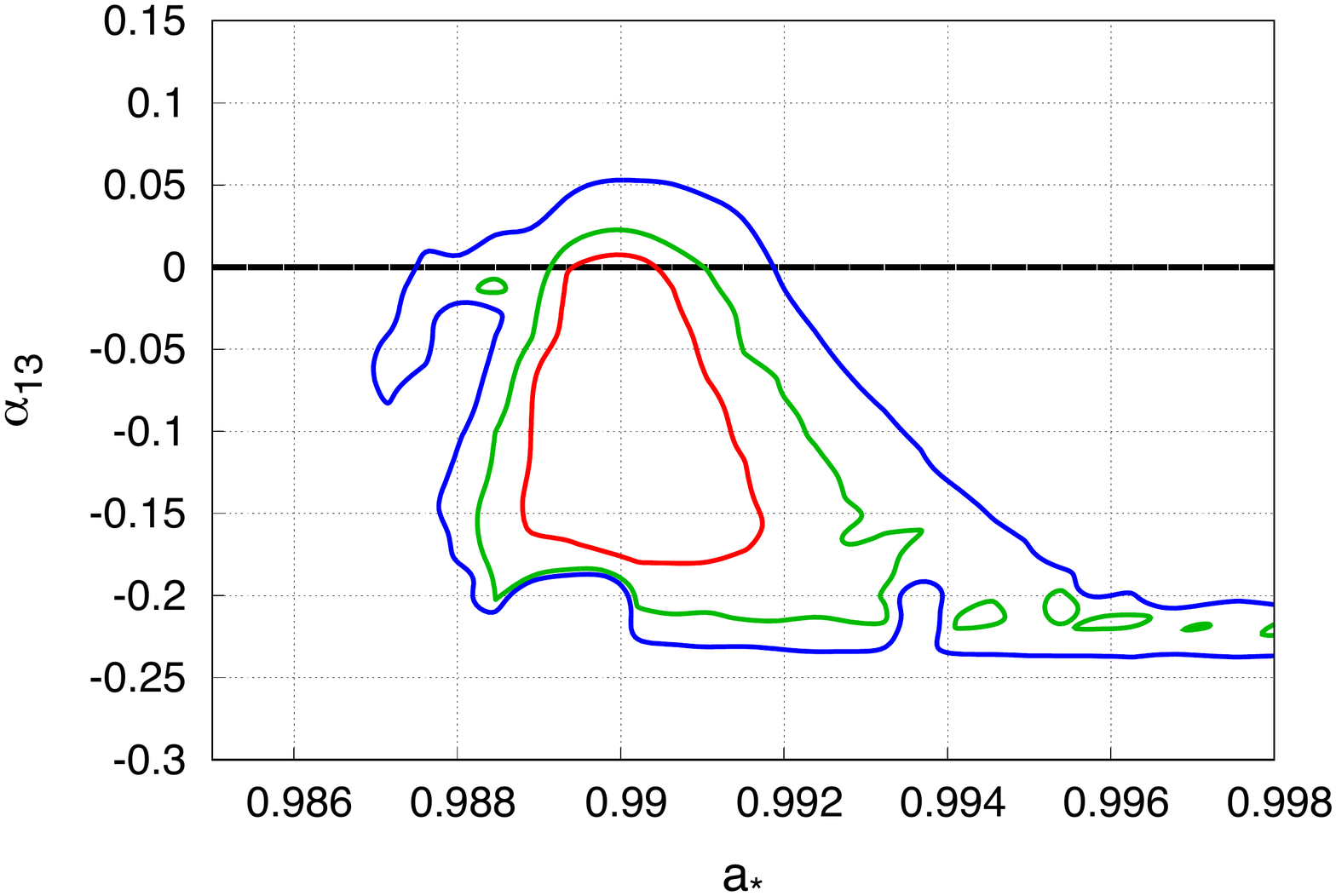}
\hspace{0.5cm}
\includegraphics[width=0.45\textwidth,trim={1.5cm 2.0cm 1.0cm 0cm},clip]{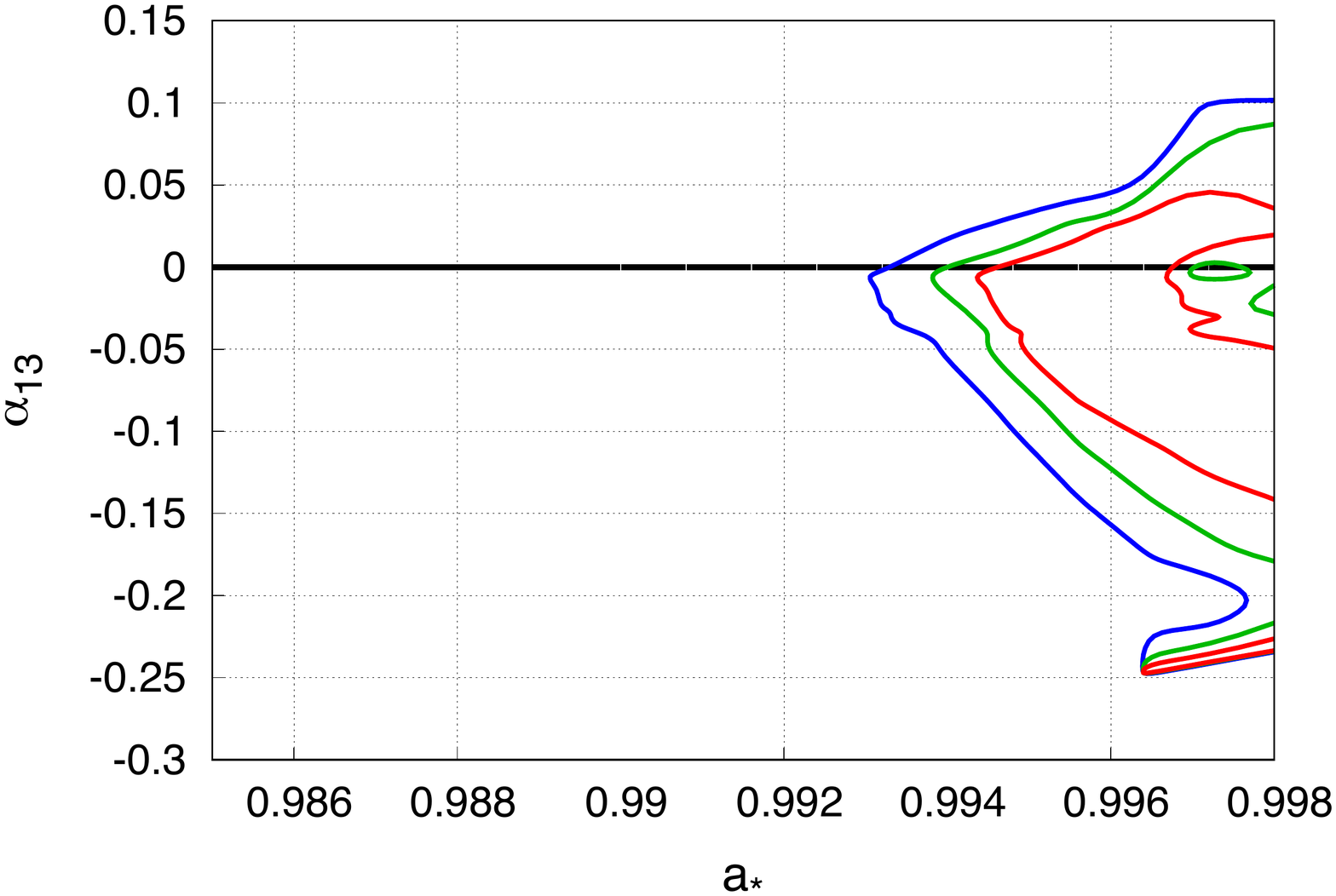}
\end{center}
\vspace{-0.5cm}
\caption{Constraints on the spin parameter $a_*$ and on the deformation parameter $\alpha_{13}$ from the 2007 \textsl{Suzaku} data of the X-ray binary GRS~1915+105. In the left panel, we analyzed the data with the current version of {\sc relxill\_nk} with an infinitesimally thin disk. In the right panel, we used the new version of {\sc relxill\_nk} with $\dot{M}/\dot{M}_{\rm Edd} = 0.2$. The red, green, and blue curves mark, respectively, the 68\%, 90\%, and 99\% confidence level contours for two relevant parameters ($\Delta\chi^2 = 2.30$, 4.61, and 9.21, respectively). The thick horizontal line marks the Kerr solution $\alpha_{13} = 0$. \label{f-grs}}
\end{figure*}

\begin{figure*}[t]
\begin{center}
\includegraphics[width=0.95\textwidth,trim={0cm 0cm 0cm 0cm},clip]{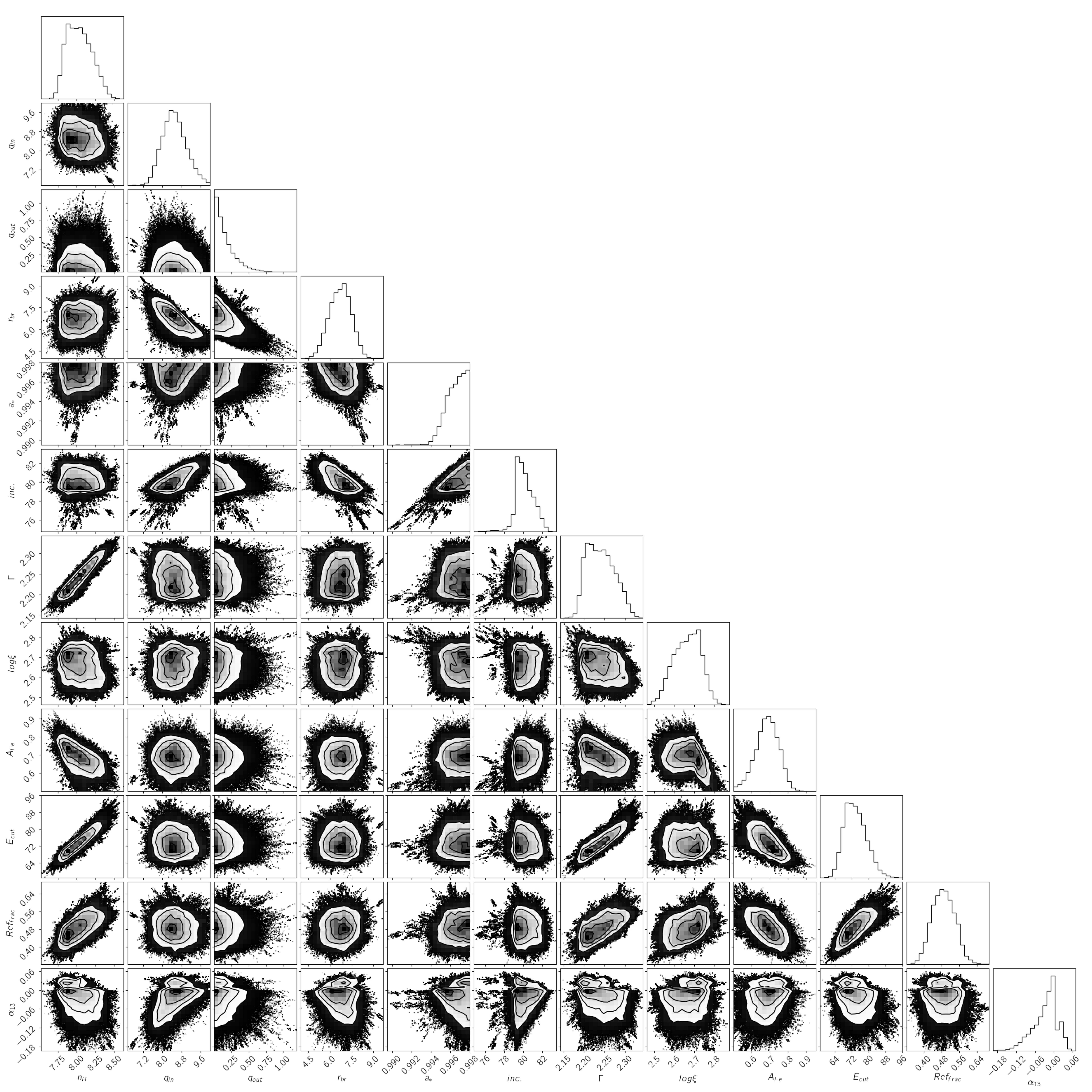}
\end{center}
\vspace{-0.3cm}
\caption{Corner plot for all the free parameter-pairs in the model with $\dot{M}/\dot{M}_{\rm Edd} = 0.2$ after the MCMC run. \label{f-mcmc}}
\vspace{0.4cm}
\end{figure*}

\begin{figure*}[t]
\begin{center}
\includegraphics[width=0.45\textwidth,trim={0cm 0cm 0cm 0cm},clip]{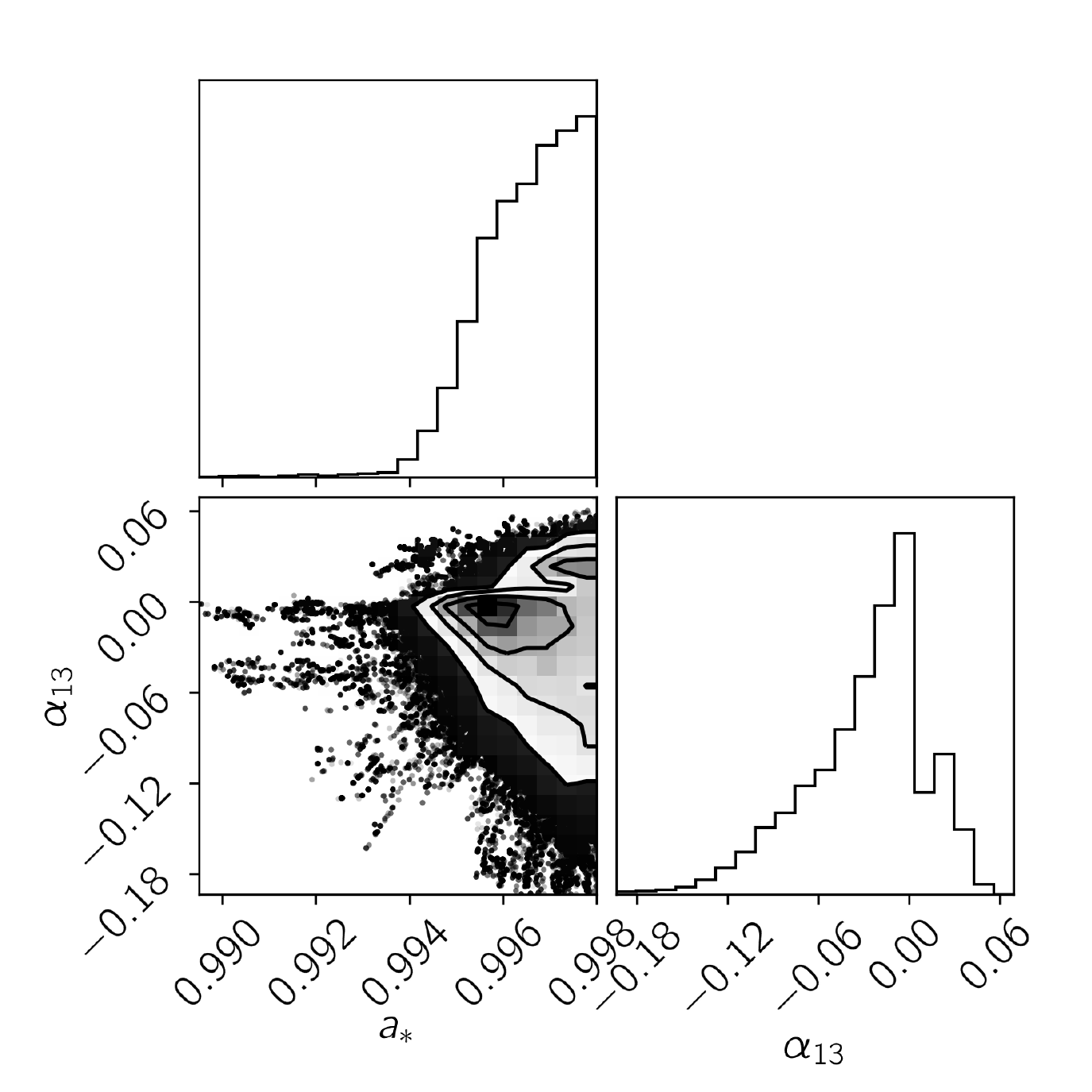}
\end{center}
\vspace{-0.3cm}
\caption{1-, 2-, and 3-$\sigma$ confidence contours for the spin parameter $a_*$ and the deformation parameter $\alpha_{13}$ in the model with $\dot{M}/\dot{M}_{\rm Edd} = 0.2$ after the MCMC run. \label{f-mcmc2}}
\vspace{0.4cm}
\end{figure*}

The analysis of the 2007 \textsl{Suzaku} observation of GRS~1915+105 reported in~\citet{2019ApJ...884..147Z} provides, at the moment, one of the most stringent and robust tests of the Kerr metric with {\sc relxill\_nk}. \textsl{Suzaku} observed GRS~1915+105 on 27~May~2007 (obs. ID~402071010), when the source was in the low-hard state, for approximately 117~ks. After all efficiencies and screening, the net exposure time is 29~ks for the XIS1 camera and 53~ks for HXD/PIN, while the other XIS units either were turned off or run in a special timing mode. As shown in \citet{2019ApJ...884..147Z}, the hardness of the source was quite stable in the 2007 \textsl{Suzaku} observation. The spectrum is clearly dominated by a strong relativistic reflection component, with a clear broad iron line around 6~keV and a Compton hump peaking around 20~keV. We do not see any thermal component from the disk, which is also welcome because the non-relativistic reflection model employed is {\sc xillver}, which should only be used for cold accretion disks. The quality of the \textsl{Suzaku} data is very good and we have both a high energy resolution near the iron line with the XIS1 instrument and a broad energy band when we add the PIN data.

As discussed in~\citet{2019ApJ...884..147Z}, the XSPEC model {\sc tbabs$\times$relxill\_nk} fits the data well and it seems that we do not need other components\footnote{Note that if we add a non-relativistic reflection component to describe some possible cooler material at larger distances, we find that its normalization would be very low and we would not improve the quality of the fit.}. {\sc tbabs} describes the Galactic absorption~\citep{2000ApJ...542..914W}, and the hydrogen column density is left free in all fits. {\sc relxill\_nk} is our relativistic reflection spectrum in the Johannsen metric with non-vanishing deformation parameter $\alpha_{13}$ and we consider two models: the infinitesimally thin disk ($\dot{M}/\dot{M}_{\rm Edd} = 0$) and the model with $\dot{M}/\dot{M}_{\rm Edd} = 0.2$. The best-fit values for the two models are reported in Tab.~\ref{t-fit}, where the parameter uncertainties correspond to the 90\% confidence level. Best-fit models and ratio plots are shown in Fig.~\ref{f-ratio}. As our interest here is in the impact of the disk thickness on tests of the Kerr metric, we also show the constraints on the black hole spin and the deformation parameter in Fig.~\ref{f-grs} after marginalization over all other free parameters.

The fit of the model with a disk of finite thickness is only a bit better, but not significantly better, than the model with an infinitesimally thin disk ($\Delta\chi^2 = 8.23$). The measurements of most model parameters are consistent; in particular, there is no difference in the final constraint on the deformation parameter $\alpha_{13}$. Both models require a very high spin parameter; the two measurements are slightly different if we believe in the statistical uncertainty of the fits, but that is indeed too low to expect that systematic uncertainties are not dominant. The model parameter presenting some difference in the two measurements is the disk inclination angle, and the model with a disk of finite thickness requires a very high value of $\iota$.

In both models, we find a very high inner emissivity index and a very low outer emissivity index. We interpret this result as a possible indication of a corona with a ring-like axisymmetry geometry located just above the accretion disk, which actually would be fitted better with a twice broken power-law with very steep emissivity profile over the inner region, then flattening in the intermediate region, and falling off approximately as $r^{-3}$ over the outer region~\citep{2003MNRAS.344L..22M,2011MNRAS.414.1269W,2015MNRAS.449..129W}. Such a coronal geometry above the accretion disk would predict the Componization of the relativistic reflection component that, when not taken into account in the XSPEC model, would lead to residuals similar to those shown in the lower panels of Fig.~\ref{f-ratio}~\citep{2015MNRAS.448..703W}. We note that other authors have interpreted very high inner emissivity indices from the fit as a deficiency of the model, in particular of the assumption of a constant ionization profile of the disk~\citep{2019MNRAS.485..239K}. Such an interpretation would presumably lead to a different measurement of the spin and of the deformation parameter~\citep{2020MNRAS.492..405S}, and we plan to leave the study of such a possibility and its impact in the constraint on $\alpha_{13}$ to future work.

Since the model {\sc tbabs$\times$relxill\_nk} has 13~free parameters in our fits and the $\chi^2$ minimizing algorithm of XSPEC has often problems to reliably find a minimum in complicated $\chi^2$ landscapes, we perform a Markov Chain Monte-Carlo (MCMC) analysis of the case with $\dot{M}/\dot{M}_{\rm Edd} = 0.2$ using the python script by Jeremy Sanders which uses {\sc emcee} (MCMC Ensemble sampler implementing Goodman \& Weare algorithm)\footnote{Available on github at \href{https://github.com/jeremysanders/xspec_emcee}{https://github.com/jeremysanders/xspec\_emcee}.}. We use 200~walkers of 10,000~iterations, after burning the initial 1,000 iterations (which is around 100~times the autocorrelation length). Fig.~\ref{f-mcmc} shows the corner plot with all the 1- and 2-dimensional projections of the posterior probability distributions of all the free parameters (we only omit the normalization of {\sc relxill\_nk}). The 2-dimensional projections report the 1-, 2-, and 3-$\sigma$ confidence level limits for two relevant parameters. In Fig.~\ref{f-mcmc2} we zoom into the spin parameter vs deformation parameter panel of Fig.~\ref{f-mcmc}. We note that the results of the MCMC analysis is consistent with the results obtained with XSPEC; in particular, the constraints in Fig.~\ref{f-mcmc2} are very similar to the constraints in the right panel in Fig.~\ref{f-grs}.

\vspace{0.3cm}


\section{Concluding remarks}\label{sec:conclusions}

The possibility of performing precision tests of general relativity in the strong gravity region around black holes using X-ray reflection spectroscopy is determined by our capability of limiting the systematic uncertainties (broadly defined) in the final measurement of possible deviations from the Kerr background. The work presented in this paper is a step of our program to develop a sufficiently sophisticated relativistic reflection model to perform precision tests of the Kerr black hole hypothesis.

In current relativistic reflection models, the accretion disk is assumed to be infinitesimally thin, while in reality it has a finite thickness, which should increase as the mass accretion rate increases. Here we have presented an extension of {\sc relxill\_nk} in which the disk has a finite thickness by implementing the disk geometry proposed in \citet{2018ApJ...855..120T}. With the current structure of the model, we cannot add the mass accretion rate as a new model parameter capable of varying over some range, as this would make the FITS file too large. We have thus constructed FITS files for specific values of the mass accretion rate of the source. With our current version of the ray-tracing code, the construction of a single FITS file for a specific value of $\dot{M}/\dot{M}_{\rm Edd}$ requires about two weeks on a computer cluster with about 250~cores. The size of the FITS file is about 1.3~GB.

In Section~\ref{sec:grs1915} we have analyzed the 2007 \textsl{Suzaku} observation of GRS~1915+105 with {\sc relxill\_nk} assuming either that the accretion disk is infinitesimally thin and that the disk has a finite thickness with $\dot{M}/\dot{M}_{\rm Edd} = 0.2$, which is the estimate inferred from the \textsl{Suzaku} observation and the known values of mass and distance of the source. Our analysis does not show significant difference in the estimate of the model parameters and, in particular, in the constraint on the deformation parameter $\alpha_{13}$. 
We should stress that we have analyzed very high-quality data: GRS~1915+105 is a bright source and \textsl{Suzaku} has both a good energy resolution near the iron line (which is the most informative part of the reflection spectrum concerning the spacetime metric) and high energy data to fit the Compton hump. The source is also characterized by a high disk inclination angle, which should maximize the impact of the thickness of the disk. As of now, the analysis of these data provides one of the most stringent constraints on the Kerr metric with {\sc relxill\_nk}, so this motivated us to use the new model with this observation. It is possible that the weak impact of the disk thickness on the analysis of this source is related to the fact that the estimate of the radiative efficiency $\eta$ is high, which makes the disk quite thin even if $\dot{M}/\dot{M}_{\rm Edd} = 0.2$. However, this is always the case for sources used to test the Kerr metric, because for low values of $\eta$ the signature of the background metric on the reflection spectrum is weak and we cannot constrain the deformation parameter due to parameter degeneracy.

Last, we note that there is no disagreement between our results and those found in \citet{2018ApJ...855..120T}, but a comparison is not straightforward. In our model the intensity profile is described a broken power-law, while \citet{2018ApJ...855..120T} consider the profile generated by a point-like lamppost corona. \citet{2018ApJ...855..120T} find that the disk thickness leads to underestimating the black hole spin parameter when the data are fitted with a infinitesimally thin disk model, but their input parameter is $a_* = 0.9$, so $\eta$ is lower and the thickness of the disk is higher. Moreover, they assume a point-like lamppost corona with height $h = 3$~$M$: for such a low value of $h$, the difference of the intensity profile between a disk of finite thickness and an infinitesimally thin disk is quite pronounced. In our case, since we have analyzed a source with high $\eta$, the thickness of the disk is lower and probably for this reason we do not see any clear modeling bias in the measurements of the model parameters. Note that the purpose of implementing a disk with finite thickness in {\sc relxill\_nk} is not primarily to fit sources with thicker disk. Our goal is to get stringent constraints on the deformation parameters and for this reason we have analyzed the \textsl{Suzaku} observation of GRS~1915+105, as it represents one of the most stringent tests of the Kerr metric.

We want to stress that the thickness of the disk is one of the current common model simplifications among many others, and presumably not the most crucial one. The present extension of {\sc relxill\_nk} implementing a disk with finite thickness does not aim to perform generic tests of the Kerr metric using X-ray reflection spectroscopy. It is meant to arrive at an estimate of the impact of the disk thickness on our capabilities of testing the Kerr black hole hypothesis. The model has still a number of simplifications that inevitably lead to modeling bias currently not well under control. Uncertainties in the coronal geometry, simplifications in the calculations of the reflection spectrum in the rest-frame of the accreting gas on the disk, impact of magnetic fields on the disk structure, etc. are all effects that need to be investigated in order to improve the capability of X-ray reflection spectroscopy to study accreting black holes.

\vspace{0.3cm}


{\bf Acknowledgments --}
This work was supported by the Innovation Program of the Shanghai Municipal Education Commission, Grant No.~2019-01-07-00-07-E00035, and the National Natural Science Foundation of China (NSFC), Grant No.~11973019.
A.B.A. also acknowledges support from the Shanghai Government Scholarship (SGS).
J.A.G. and S.N. acknowledge support from the Alexander von Humboldt Foundation.
C.B., J.A.G., S.N., and A.T. are members of the International Team~458 at the International Space Science Institute (ISSI), Bern, Switzerland, and acknowledge support from ISSI during the meetings in Bern.


\appendix

\section{Johannsen metric}\label{app:metric}

In Boyer-Lindquist-like coordinates, the line element of the Johannsen metric reads~\citep{jj}
\be\label{eq-jm}
ds^2 &=&-\frac{\tilde{\Sigma}\left(\Delta-a^2A_2^2\sin^2\theta\right)}{B^2}dt^2
+\frac{\tilde{\Sigma}}{\Delta}dr^2+\tilde{\Sigma} d\theta^2 
-\frac{2a\left[\left(r^2+a^2\right)A_1A_2-\Delta\right]\tilde{\Sigma}\sin^2\theta}{B^2}dtd\phi \nonumber\\
&&+\frac{\left[\left(r^2+a^2\right)^2A_1^2-a^2\Delta\sin^2\theta\right]\tilde{\Sigma}\sin^2\theta}{B^2}d\phi^2
\ee
where $M$ is the black hole mass, $a = J/M$, $J$ is the black hole spin angular momentum, $\tilde{\Sigma} = \Sigma = f$, and
\be
\Sigma = r^2 + a^2 \cos^2\theta \, , \qquad
\Delta = r^2 - 2 M r + a^2 \, , \qquad
B = \left(r^2+a^2\right)A_1-a^2A_2\sin^2\theta \, .
\ee
The functions $f$, $A_1$, $A_2$, and $A_5$ are defined as
\be
f = \sum^\infty_{n=3} \epsilon_n \frac{M^n}{r^{n-2}} \, , \quad
A_1 = 1 + \sum^\infty_{n=3} \alpha_{1n} \left(\frac{M}{r}\right)^n \, , \quad
A_2 = 1 + \sum^\infty_{n=2} \alpha_{2n}\left(\frac{M}{r}\right)^n \, , \quad
A_5 = 1 + \sum^\infty_{n=2} \alpha_{5n}\left(\frac{M}{r}\right)^n \, ,
\ee
where $\{ \epsilon_n \}$, $\{ \alpha_{1n} \}$, $\{ \alpha_{2n} \}$, and $\{ \alpha_{5n} \}$ are four infinite sets of deformation parameters without constraints from the Newtonian limit and Solar System experiments. The leading order deformation parameters are thus $\epsilon_3$, $\alpha_{13}$, $\alpha_{22}$, and $\alpha_{52}$. In this paper, we have only considered the deformation parameter $\alpha_{13}$ because it has the strongest impact on the shape of the reflection spectrum, but all our results can be easily extended to metrics with other non-vanishing deformation parameters as well as, more in general, to any stationary, axisymmetric, and asymptotically flat black hole spacetime.

In order to avoid spacetimes with pathological properties, we must impose some constraints on the values of $a_*$ and $\alpha_{13}$. As in the case of the Kerr spacetime, we must impose that $| a_* | \le 1$; for $| a_* | > 1$ there is no event horizon and the solution describes the spacetime of a naked singularity. As discussed in \citet{564}, we have to impose the following constraint on $\alpha_{13}$
\be
\label{eq-constraints}
\alpha_{13} > - \frac{1}{2} \left( 1 + \sqrt{1 - a^2_*} \right)^4 \, .
\ee



\begin{thebibliography}{99}

\bibitem[Abdikamalov et al.(2019)]{2019ApJ...878...91A} Abdikamalov, A.~B., Ayzenberg, D., Bambi, C., et al.\ 2019, \apj, 878, 91

\bibitem[Ayzenberg \& Yunes(2018)]{2018CQGra..35w5002A} Ayzenberg, D., \& Yunes, N.\ 2018, Classical and Quantum Gravity, 35, 235002

\bibitem[Bambi(2017)]{review} Bambi, C.\ 2017, Reviews of Modern Physics, 89, 025001

\bibitem[Bambi(2018)]{d2} Bambi, C.\ 2018, Annalen der Physik, 530, 1700430

\bibitem[Bambi et al.(2017)]{2017ApJ...842...76B} Bambi, C., C{\'a}rdenas-Avenda{\~n}o, A., Dauser, T., et al.\ 2017, \apj, 842, 76

\bibitem[Bambi et al.(2014)]{d1} Bambi, C., Malafarina, D., \& Tsukamoto, N.\ 2014, \prd, 89, 127302

\bibitem[Bardeen et al.(1972)]{1972ApJ...178..347B} Bardeen, J.~M., Press, W.~H., \& Teukolsky, S.~A.\ 1972, \apj, 178, 347

\bibitem[Blum et al.(2009)]{2009ApJ...706...60B} Blum, J.~L., Miller, J.~M., Fabian, A.~C., et al.\ 2009, \apj, 706, 60

\bibitem[Brenneman \& Reynolds(2006)]{2006ApJ...652.1028B} Brenneman, L.~W., \& Reynolds, C.~S.\ 2006, \apj, 652, 1028

\bibitem[Cao et al.(2018)]{2018PhRvL.120e1101C} Cao, Z., Nampalliwar, S., Bambi, C., et al.\ 2018, \prl, 120, 051101

\bibitem[Carballo-Rubio et al.(2020)]{e5} Carballo-Rubio, R., Di Filippo, F., Liberati, S., et al.\ 2020, \prd, 101, 084047 

\bibitem[Carter(1971)]{k2} Carter, B.\ 1971, \prl, 26, 331

\bibitem[Chru{\'s}ciel et al.(2012)]{k4} Chru{\'s}ciel, P.~T., Costa, J.~L., \& Heusler, M.\ 2012, Living Reviews in Relativity, 15, 7

\bibitem[Cunningham(1975)]{1975ApJ...202..788C} Cunningham, C.~T.\ 1975, \apj, 202, 788

\bibitem[Dauser et al.(2010)]{2010MNRAS.409.1534D} Dauser, T., Wilms, J., Reynolds, C.~S., et al.\ 2010, \mnras, 409, 1534

\bibitem[Dauser et al.(2013)]{2013MNRAS.430.1694D} Dauser, T., Garcia, J., Wilms, J., et al.\ 2013, \mnras, 430, 1694

\bibitem[Dvali \& Gomez(2011)]{e3} Dvali, G., \& Gomez, C.\ 2013, Fortschritte der Physik, 61, 742

\bibitem[Fabian et al.(1989)]{1989MNRAS.238..729F} Fabian, A.~C., Rees, M.~J., Stella, L., et al.\ 1989, \mnras, 238, 729

\bibitem[Fender \& Belloni(2004)]{2004ARA&A..42..317F} Fender, R., \& Belloni, T.\ 2004, \araa, 42, 317

\bibitem[Garc{\'\i}a et al.(2014)]{2014ApJ...782...76G} Garc{\'\i}a, J., Dauser, T., Lohfink, A., et al.\ 2014, \apj, 782, 76

\bibitem[Garc{\'\i}a et al.(2013)]{2013ApJ...768..146G} Garc{\'\i}a, J., Dauser, T., Reynolds, C.~S., et al.\ 2013, \apj, 768, 146

\bibitem[George \& Fabian(1991)]{1991MNRAS.249..352G} George, I.~M., \& Fabian, A.~C.\ 1991, \mnras, 249, 352

\bibitem[Giddings \& Psaltis(2016)]{e4} Giddings, S.~B., \& Psaltis, D.\ 2018, \prd, 97, 084035

\bibitem[Gott et al.(2019)]{2019CQGra..36e5007G} Gott, H., Ayzenberg, D., Yunes, N., et al.\ 2019, Classical and Quantum Gravity, 36, 055007

\bibitem[Greiner et al.(2001)]{2001A&A...373L..37G} Greiner, J., Cuby, J.~G., McCaughrean, M.~J., et al.\ 2001, \aap, 373, L37

\bibitem[Johannsen(2013)]{jj} Johannsen, T.\ 2013, \prd, 88, 044002

\bibitem[Johannsen(2016)]{2016CQGra..33l4001J} Johannsen, T.\ 2016, Classical and Quantum Gravity, 33, 124001

\bibitem[Kammoun et al.(2019)]{2019MNRAS.485..239K} Kammoun, E.~S., Dom{\v{c}}ek, V., Svoboda, J., et al.\ 2019, \mnras, 485, 239

\bibitem[Kerr(1963)]{k1} Kerr, R.~P.\ 1963, \prl, 11, 237

\bibitem[Krawczynski(2018)]{2018GReGr..50..100K} Krawczynski, H.\ 2018, General Relativity and Gravitation, 50, 100

\bibitem[Lindquist(1966)]{1966AnPhy..37..487L} Lindquist, R.~W.\ 1966, Annals of Physics, 37, 487

\bibitem[Liu et al.(2019)]{2019PhRvD..99l3007L} Liu, H., Abdikamalov, A.~B., Ayzenberg, D., et al.\ 2019, \prd, 99, 123007

\bibitem[McClintock et al.(2006)]{2006ApJ...652..518M} McClintock, J.~E., Shafee, R., Narayan, R., et al.\ 2006, \apj, 652, 518

\bibitem[Miniutti et al.(2003)]{2003MNRAS.344L..22M} Miniutti, G., Fabian, A.~C., Goyder, R., et al.\ 2003, \mnras, 344, L22

\bibitem[Nampalliwar et al.(2019)]{2019arXiv190312119N} Nampalliwar, S., Xin, S., Srivastava, S., et al.\ 2019, arXiv e-prints, arXiv:1903.12119

\bibitem[Novikov \& Thorne(1973)]{1973blho.conf..343N} Novikov, I.~D., \& Thorne, K.~S.\ 1973, Black Holes (les Astres Occlus), 343

\bibitem[Page \& Thorne(1974)]{1974ApJ...191..499P} Page, D.~N., \& Thorne, K.~S.\ 1974, \apj, 191, 499

\bibitem[Penna et al.(2010)]{2010MNRAS.408..752P} Penna, R.~F., McKinney, J.~C., Narayan, R., et al.\ 2010, \mnras, 408, 752

\bibitem[Psaltis \& Johannsen(2012)]{2012ApJ...745....1P} Psaltis, D., \& Johannsen, T.\ 2012, \apj, 745, 1

\bibitem[Reid et al.(2014)]{2014ApJ...796....2R} Reid, M.~J., McClintock, J.~E., Steiner, J.~F., et al.\ 2014, \apj, 796, 2

\bibitem[Reynolds(2014)]{2014SSRv..183..277R} Reynolds, C.~S.\ 2014, \ssr, 183, 277

\bibitem[Riaz et al.(2020a)]{2020MNRAS.491..417R} Riaz, S., Ayzenberg, D., Bambi, C., et al.\ 2020a, \mnras, 491, 417

\bibitem[Riaz et al.(2020b)]{2019arXiv191106605R} Riaz, S., Ayzenberg, D., Bambi, C., et al.\ 2020b, \apj, 895, 61 

\bibitem[Robinson(1975)]{k3} Robinson, D.~C.\ 1975, \prl, 34, 905

\bibitem[Ross \& Fabian(2005)]{2005MNRAS.358..211R} Ross, R.~R., \& Fabian, A.~C.\ 2005, \mnras, 358, 211

\bibitem[Shakura \& Sunyaev(1973)]{1973A&A....24..337S} Shakura, N.~I., \& Sunyaev, R.~A.\ 1973, \aap, 500, 33 

\bibitem[Shreeram \& Ingram(2020)]{2020MNRAS.492..405S} Shreeram, S., \& Ingram, A.\ 2020, \mnras, 492, 405 

\bibitem[Speith et al.(1995)]{1995CoPhC..88..109S} Speith, R., Riffert, H., \& Ruder, H.\ 1995, Computer Physics Communications, 88, 109

\bibitem[Steiner et al.(2010)]{2010ApJ...718L.117S} Steiner, J.~F., McClintock, J.~E., Remillard, R.~A., et al.\ 2010, \apjl, 718, L117

\bibitem[Sunyaev \& Truemper(1979)]{1979Natur.279..506S} Sunyaev, R.~A., \& Truemper, J.\ 1979, \nat, 279, 506

\bibitem[Taylor \& Reynolds(2018a)]{2018ApJ...855..120T} Taylor, C., \& Reynolds, C.~S.\ 2018a, \apj, 855, 120

\bibitem[Taylor \& Reynolds(2018b)]{2018ApJ...868..109T} Taylor, C., \& Reynolds, C.~S.\ 2018b, \apj, 868, 109

\bibitem[Tripathi et al.(2018)]{564} Tripathi, A., Nampalliwar, S., Abdikamalov, A.~B., et al.\ 2018, \prd, 98, 023018

\bibitem[Tripathi et al.(2019a)]{2019ApJ...875...56T} Tripathi, A., Nampalliwar, S., Abdikamalov, A.~B., et al.\ 2019a, \apj, 875, 56

\bibitem[Tripathi et al.(2019b)]{2019ApJ...874..135T} Tripathi, A., Yan, J., Yang, Y., et al.\ 2019b, \apj, 874, 135

\bibitem[Tripathi et al.(2020)]{2019arXiv191203868T} Tripathi, A., Zhou, B., Abdikamalov, A.~B., et al.\ 2020, \prd, 101, 064030 

\bibitem[Wilkins \& Fabian(2011)]{2011MNRAS.414.1269W} Wilkins, D.~R., \& Fabian, A.~C.\ 2011, \mnras, 414, 1269

\bibitem[Wilkins \& Gallo(2015a)]{2015MNRAS.448..703W} Wilkins, D.~R., \& Gallo, L.~C.\ 2015a, \mnras, 448, 703

\bibitem[Wilkins \& Gallo(2015b)]{2015MNRAS.449..129W} Wilkins, D.~R., \& Gallo, L.~C.\ 2015b, \mnras, 449, 129

\bibitem[Will(2014)]{2014LRR....17....4W} Will, C.~M.\ 2014, Living Reviews in Relativity, 17, 4

\bibitem[Wilms et al.(2000)]{2000ApJ...542..914W} Wilms, J., Allen, A., \& McCray, R.\ 2000, \apj, 542, 914

\bibitem[Zhang et al.(1997)]{1997ApJ...482L.155Z} Zhang, S.~N., Cui, W., \& Chen, W.\ 1997, \apjl, 482, L155

\bibitem[Zhang et al.(2019)]{2019ApJ...884..147Z} Zhang, Y., Abdikamalov, A.~B., Ayzenberg, D., et al.\ 2019, \apj, 884, 147

\bibitem[Zhou et al.(2019)]{2019PhRvD..99j4031Z} Zhou, M., Abdikamalov, A.~B., Ayzenberg, D., et al.\ 2019b, \prd, 99, 104031

\bibitem[Zhou et al.(2020a)]{2019arXiv190805177Z} Zhou, B., Tripathi, A., Abdikamalov, A.~B., et al.\ 2020a, European Physical Journal C, 80, 400 

\bibitem[Zhou et al.(2020b)]{2020PhRvD.101d3010Z} Zhou, M., Ayzenberg, D., Bambi, C., et al.\ 2020b, \prd, 101, 043010

\bibitem[Zhou et al.(2018)]{2018PhRvD..98b4007Z} Zhou, M., Cao, Z., Abdikamalov, A., et al.\ 2018, \prd, 98, 024007

\end{thebibliography}
\end{document}